\documentclass[aps,prl,reprint,groupedaddress,amssymb]{revtex4-2}

\usepackage{amssymb}% Adds mathematical symbols
\usepackage{graphicx}% Includes figure files
\usepackage{dcolumn}% Align table columns on decimal point

\usepackage{xcolor}

\begin{document}

%\preprint{\textcopyright\,NPL 2020}

\title{Theory of cell membrane interaction with glass}


\author{Richard W. Clarke}
%\altaffiliation{University of Cambridge, U.K.}
%\email[]{richard.w.clarke@npl.co.uk; \textcopyright\,NPL 2020}
\email[]{richard.w.clarke@npl.co.uk}
%\thanks{}
\affiliation{National Physical Laboratory, Teddington, TW11 0LW, U.K.}


\date{January 19, 2021}
%\date{\today}

\begin{abstract}
There are three regimes of cell membrane interaction with glass - Tight and loose adhesion, separated by repulsion. Explicitly including hydration, this paper evaluates the pressure between the surfaces as functions of distance for ion-correlation and ion-screened electrostatics, and electromagnetic fluctuations. The results agree with data for tight adhesion energy (0.5-3 vs 0.4-4 mJ/m\textsuperscript{2}), detachment pressure (7.9 vs 9 MPa), and peak repulsion (3.4-7.5 vs 5-10 kPa), also matching the repulsion's distance dependence upon renormalization by steric pressure mainly from undulations.
\end{abstract}
\maketitle
\begin{figure}
\includegraphics{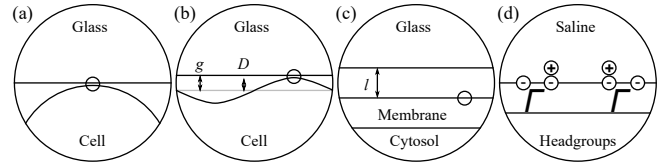}
\caption{\label{fig:diagrams}Glass-membrane interaction across saline at progressive magnifications of the circled regions. (a) Glass near cell. (b) Undulations extend the interaction range: Peak depth \textit{D} is slightly less than mean gap \textit{g}. (c) Approach length \textit{l} indexes strong variation in surface pressures. (d) Most net surface charge is from sialic acids (-) at the edge of the membrane's external hydration layer, on ganglioside headgroups ($\Gamma$) embedded in it. Divalent cations (bold +) like Ca\textsuperscript{2+} can bridge charges on the two surfaces.}
\end{figure}

\textit{Introduction.} Physical interactions between cell membranes and with their substrates are central to biology. Crucial to studying them is the refinement of modelling for lipid bilayer, the only common component of cell membranes in general, and the most important. I show here that membrane-saline-glass is an even better system for characterizing cell membrane biophysics than previously recognized, by resolving theoretically the interaction pressure between lipid bilayer and glass, which unusually and yet usefully exhibits repulsion between two zones of attraction. Tight \cite{Hamill1981} and loose \cite{Almers1983} adhesion of cell membrane to glass pipets respectively facilitate studies of ion channels' electrophysiology and spatial distributions. The tight adhesion energy is vastly stronger; two laboratories both place it in the range 0.4-4.0 mJ/m\textsuperscript{2} \cite{Opsahl1994,Ursell2011}. At intermediate separations, glass instead repelling membrane allows nanopipets to map the topography \cite{Hansma1989} and stiffness \cite{Clarke2016} of individual cells. The peak repulsion is in the range 5-10 kPa \cite{Ursell2011,Yang2013}, though measurements are as yet only incidental to the need for electrophysiological protocol to achieve tight seals while minimising perturbative membrane tension \cite{Hamill1997,Ursell2011,Sachs2015}. To reconcile these regimes of interaction by calculating the pressure between cell membrane and glass across saline, this paper has to make various theoretical advances: It incorporates hydration rationally, estimates the full distance dependence of the pressure from electromagnetic fluctuations, and develops a model for ion-correlated attraction. Results agree with many classes of measurement, and hence among other predictions the theory clarifies the views that tight adhesion is to lipid bilayer \cite{Opsahl1994}, but includes hydration \cite{Hamill2014}, and glycans \cite{Bae2011}, though short ones as illustrated in Fig.~\ref{fig:diagrams}.

\begin{figure}
\includegraphics{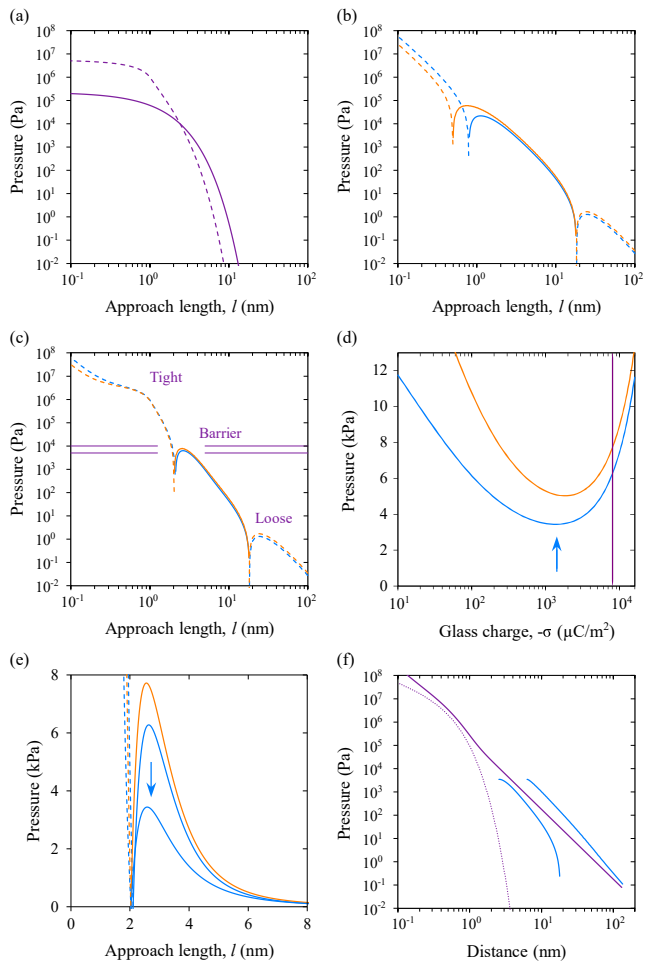}
\caption{\label{fig:graphs}Calculated pressures between lipid bilayer and silicate (blue), borosilicate (orange), or either (purple). Dashes indicate attraction. (a) Electrostatic pressures from ion-screened double-layer repulsion and ion-correlated attraction. (b) Dispersion pressures. (c) Sum of electrostatic and dispersion pressures reproduces the observed 5-10 kPa peak barrier pressure (rules) between loose and tight adhesion. (d) Peak repulsion versus glass surface charge density, usually -8.01 mC/m\textsuperscript{2} (line). At -1.42 mC/m\textsuperscript{2} (arrow) the peak repulsion for silicate is minimized to 3.4 kPa. (e) Peak repulsion and its minimization for silicate. (f) Renormalization (right) of the pressure barrier for silicate by the steric pressure, mainly from membrane undulation rather than lipid protrusion (dots). To avoid curve overlap for clarity, the minimized barrier is shown.}
\end{figure}

The experimental characterization of the system is best summarized in terms of protocols for tight adhesion of a membrane patch to a glass pipet. With a nanopipet the tight seal forms on the $\approx$100 nm diameter tip face \cite{Novak2013} whereas micropipets usually seal to blebs extruded inwards \cite{Hamill1981}. Both adhesion geometries are called a gigaseal as they ideally have 10-100 G$\Omega$ resistance. Forming such a seal requires clean, smooth glass and a clear cell membrane \cite{Ogden1994} so patch-clamp studies either target (blebs on) new neurons without perineuronal nets, or treat differentiated cells with proteases. In addition, positive hydrostatic pressure, typically 3-5 kPa \cite{Yang2013}, is applied to the pipet during its approach, producing a flow profile strong enough to clear membrane proteins \cite{Jonsson2012} from an exclusion zone where the seal forms \cite{Suchyna2009} but not necessarily the central flow stagnation point, sometimes leaving an individual ion channel. The flow also pushes the area downwards \cite{Rheinlaender2013}. Thus, at the next stage when pressure is suddenly released or reversed, the elastic energy stored in the cytoskeleton \cite{SachsSivaselvan2015} springs the relatively clear patch of membrane up to the pipet, where its proximity will continue to exclude membrane proteins. To minimise the absolute tension the membrane experiences, electrophysiologists can derive approximately half the pressure required for tight adhesion from this spring upward and the remainder from subsequent suction of 2-2.5 kPa \cite{Ogden1994,Opsahl1994} or its sudden release \cite{Hamill1981,Clarke2013}. This agrees with the low end of 5-10 kPa automated patching suction \cite{Yang2013}, indicated in Fig.~\ref{fig:graphs} as the experimental range of the peak barrier pressure.  

From a theoretical perspective the pressure has separate contributions from electromagnetic fluctuations and quasi-static ionic charge displacement. This is because ions free to move in solution exponentially reduce electrostatic interaction with distance but have too much inertia to screen any but the longest-lived fluctuations \cite{MahantyNinham1976}. The electrostatic pressure is from double-layer overlap and ion-correlation, while the dispersion pressure is caused by electromagnetic fluctuations' variation with surface separation \cite{Parsegian2005}. These three components are evaluated here as functions of separation for neuron lipid bilayer approached by silicate or borosilicate glass across saline at 25$^{\circ}$C. I then renormalize the barrier section of the interaction by the steric pressure from lipid protrusion \cite{Aniansson1978} and membrane undulation \cite{Helfrich1978}.
%\newline

\textit{Electrostatic pressures.} The surface charge density of glass $\sigma_{g}$ is typically one electron per 20 nm\textsuperscript{2} or -8.01 mC/m\textsuperscript{2} \cite{Jednacak1974}. Neuronal cell membrane has a similar surface charge density $\sigma_{l}$, calculable from mass-fraction data \cite{Breckenridge1972,Drin2014}: With $\Gamma=1.57\times10^{18}$ lipids/m\textsuperscript{2}, and 50\% ionization as found experimentally for bilayers with a similar proportion of 10\% net charged lipids \cite{Israelachvili2011} Table~\ref{tab:table1} totals 0.0388 electrons per lipid in the outer leaflet, which is one electron per 16.4 nm\textsuperscript{2}, or -9.78 mC/m\textsuperscript{2}. This is mainly from the 2.2 sialic acids per ganglioside as 80\% of these are on the outer leaflet \cite{Drin2014}. Extensive hydrogen bonding would stabilize gangliosides in the membrane's hydration layer with the charged groups just beyond. Thus in this model zero approach length retains headgroup hydration as shown in Fig.~\ref{fig:diagrams}. With four water molecules around each headgroup, a volume $v\approx0.12\text{ nm\textsuperscript{3}}$, the hard wall thickness of the hydration layer is $4v\Gamma\approx0.75\text{ nm}$ \cite{Israelachvili2011}.
\begin{table}%[b]
\caption{\label{tab:table1}Neuronal bilayer lipids with net charge.}
\begin{ruledtabular}
\begin{tabular}{lcccc}
 & Bilayer & Ratio & Half net & Charge per \\
 & mol \% & in:out & charge & outer lipid \\
\colrule
Ganglioside          & 3.29 & 20:80 & -1.1 & -0.02897\\
Phosphatidylserine   & 9.47 & 90:10 & -0.5 & -0.00474\\
Cardiolipin          & 0.47 & 50:50 & -1.0 & -0.00237\\
Phosphatidylinositol & 2.55 & 85:15 & -0.5 & -0.00191\\
Phosphatidic acid    & 0.34 & 50:50 & -0.5 & -0.00085\\
\end{tabular}
\end{ruledtabular}
\end{table}
With the above surface charge densities, Poisson-Boltzmann theory gives the ion-screened electrostatic pressure from double-layer overlap as
\begin{equation}
P\approx(2\sigma_{l}\sigma_{g}/\epsilon_{r}\epsilon_{0})\text{e}^{-\kappa l}
\end{equation}
where the Debye length $1/\kappa$ is given by $\kappa^{2}=\sum_{i}\rho_{i}z_{i}^{2}e^{2}/\epsilon_{r}\epsilon_{0}k_{B}T$ with $\rho_{i}$ and $z_{i}e$ the volume density and charge of ion type \textit{i}. Approximating physiological saline as 154 mM sodium chloride \cite{Li2016} of relative permittivity 78.4 at 298.15 K gives $1/\kappa \approx$ 0.775 nm. Though there are more buffer components, and notably always approximately 1 mM calcium salts, it is not worth calculating $\kappa$ more precisely because tight adhesion is otherwise broadly independent of saline formulation \cite{Tebaykin2018}.

Although both surfaces are negatively charged, ion correlation at nanometre separations breaks the mean-field assumption underlying the derivation of the double layer pressure and generates strong electrostatic attraction \cite{Lipowsky1995,Oosawa1968,Kekicheff1993}. As tight adhesion is promoted by Ca\textsuperscript{2+} \cite{Priel2007}, ion clustering dominates in non-dilute electrolytes \cite{FranceLanord2019}, and divalent ions contribute most to potential minimization, I model ion correlation here as two stages of Ca\textsuperscript{2+} displacement concomitant with sialate rearrangement: Firstly, sialate pairing with silanolate via an intermediate Ca\textsuperscript{2+}, diameter $s\approx$ 0.51 nm (averaging over 0.1-10 ns hydration-dehydration exchange \cite{Israelachvili2011}). The cation position is electrostatically bi-stable with a typical charge-center separation to one of the anions also $\approx s$. Thus, for steric displacement of any second Ca\textsuperscript{2+} from the opposite anion (to equal average reservoir potential), and for sialate rearrangement to occur, the prospective minimized charge-center separation to the other anion, $l-s$, must approach \textit{s}. This motivates an atomic granularity factor $\text{e}^{-(l-s)/s}$ in the energy per unit area of
\begin{equation}
G(l)\approx -\frac{2e\sigma\text{e}^{-(l-s)/s}}{4\pi\epsilon_{r}\epsilon_{0}(l-s)}
\end{equation}
Here \textit{e} is electron charge magnitude, and $\sigma=\sqrt{\sigma_{l}\sigma_{g}}$. The pressure $-dG/dl$, negative for attraction, is then
\begin{equation}
P\approx -\frac{e\sigma l\text{e}^{-(l-s)/s}}{2\pi\epsilon_{r}\epsilon_{0}s(l-s)^{2}}
\end{equation}
This estimate of the ion correlated attraction from calcium bridges, plotted in Fig.~\ref{fig:graphs}, exceeds the ion screened repulsion of the two surfaces at approximately $l\approx2.4\text{ nm}$, and applies while $l\ge2s\approx1\text{ nm}$. In the second stage at tighter constrictions, the rearrangement to direct electrostatic complementarity \cite{Leckband2001} between silanolates and protruding positive headgroups will displace each sialate-Ca\textsuperscript{2+} to instead attract its glass image charge. Most hydration would remain though, as discussed in the dispersion section, precluding ion desolvation. As the ion correlation pressure cannot therefore change magnitude, for $l\le2s$ I estimate it, along with inter-surface hydrogen bonding, by the linear continuation
\begin{equation}
P=-\frac{e\sigma}{2\pi\epsilon_{r}\epsilon_{0}\text{e}}\left(\frac{12}{s^{2}}-\frac{5l}{s^{3}}\right).
\end{equation}
%\newline

\textit{Dispersion pressure.} A major success of thermal quantum field theory \cite{Casimir1948,Matsubara1955,Dzyaloshinskii1961}, the dispersion interaction energy per unit area is usually written as \cite{Israelachvili2011}
\begin{equation}
G(l)=-H_{\text{LMU}}/12\pi l^{2}
\end{equation}
Here $H_{\text{LMU}}$ is the separation-dependent Hamaker coefficient for lower and upper regions L and U separated by medium M, a sum over discrete electromagnetic fluctuation lifetimes $\nu_{n}^{-1}$ at $2\pi i n k_{B}T/\hbar=2\pi i\nu_{n}=i\xi_{n}$ \cite{Parsegian2005}: 
\begin{equation}
H_{\text{LMU}}=\frac{3}{2}k_{B}T\sum\nolimits_{n\ge0}^{\prime} \mathbb{E}_{\text{LM}} \mathbb{E}_{\text{UM}} R_{n}
\end{equation}
In this equation, which assumes uniform permeability, $\mathbb{E}_{\text{AB}}=(\epsilon_{\text{A}}-\epsilon_{\text{B}})/(\epsilon_{\text{A}}+\epsilon_{\text{B}})$ with permittivities $\epsilon$ evaluated at $i\xi_{n}$. For $n\ge1$, relativistic desynchronization is estimated by the factor $R_{n}=(1+r_{n})\text{e}^{-r_{n}}$ where $r_{n\ge1}=(2l\xi_{n}\sqrt{\epsilon_{M}})/c$
compares the return-path distance to how far light can travel across the medium during the fluctuation. These terms correspond to electromagnetic fluctuations too brief for ion movement to screen \cite{MahantyNinham1976}: The plasma frequency for Na\textsuperscript{+} and Cl\textsuperscript{-}, $\frac{1}{2\pi}\sqrt{\rho e^{2}/\epsilon_{r}\epsilon_{0}m^{*}} \approx 6.1\times10^{10} \text{ Hz}$ at 154 mM is already much lower than $\nu_{1}=3.9\times10^{13} \text{ Hz}$ and would be further decreased by ion solvation. The ions do move fast enough to screen the longer-lasting fluctuations of the $n=0$ term though, with full screening by a few nanometres as the Debye length is so short. Nevertheless, close to contact the $n=0$ term exceeds all the others combined, making it necessary to include the effect of fluctuations of higher order $q$ in its calculation \cite{Parsegian2005,Dzyaloshinskii1961}:
\begin{equation}
\frac{1}{2}\sum\nolimits_{q=1}^{\infty}[(\mathbb{E}_{\text{LM}} \mathbb{E}_{\text{UM}})^{q}/q^{3}](1+r_{0}q) \text{e}^{-r_{0}q}.
\end{equation}
Ionic screening is represented here by a factor of $R_{0}$ with $r_{0}=2\kappa l$ \cite{Parsegian2005} multiplied by $q$ to account for greater ionic screening of higher-order zero-frequency fluctuations. The term is halved to avoid double-counting it, as indicated by the dash on the main sum which would otherwise use absolute $n$ over all integers. Summing to $q\le5$ is sufficient in practice \cite{Parsegian2005}. The surface pressure $-dG/dl$, with attraction negative, is then
\begin{widetext}
\begin{equation}\label{eqn:DisP}
P=-\frac{k_{B}T}{8\pi l^{3}}
\left\{\left[\frac{1}{2}\sum\nolimits_{q\ge1}^{5}[(\mathbb{E}_{\text{LM}} \mathbb{E}_{\text{UM}})^{q}/q^{3}](r_{0}^{2}q^{2}+2r_{0}q+2)\text{e}^{-r_{0}q}\right]+\left[\sum\nolimits_{n\ge1}^{N}\mathbb{E}_{\text{LM}} \mathbb{E}_{\text{UM}} (r^{2}_{n}+2r_{n}+2)\text{e}^{-r_{n}}\right]\right\}
\end{equation}
\end{widetext}
converging by $N\approx1000$, where now L indicates the lipid headgroup layer, M the saline medium, and U the glass. A general form for permittivities' dependence on fluctuation lifetime is \cite{Israelachvili2011}:
\begin{equation}
\epsilon_{r}(i\nu)\approx1+\frac{\varepsilon_{r}-n^{2}}{1+(\nu/\nu_{r})}+\frac{n^{2}-1}{1+(\nu/\nu_{e})^{2}}
\end{equation}
For silicate glass, the low-frequency relative permittivity $\varepsilon_{r}=3.8$, the refractive index in the visible spectrum, $n=1.448$, the rotational relaxation frequency $\nu_{r}\approx10^{12} \text{ Hz}$, and its electronic transitions mainly absorb ultraviolet light at $\nu_{e}\approx3.2\times10^{15} \text{ Hz}$ \cite{Israelachvili2011}. For borosilicate, industry data \cite{GlassIndustryData} averages to low-frequency relative permittivity $\varepsilon_{r}=4.90$ and optical refractive index $n=1.480$.   

As thermal undulations continuously juxtapose sections of the membrane with the glass even at much wider mean separations \cite{Helfrich1978}, I represent the membrane permittivity by that of the outer headgroups, including their hydration. This is consistent with measuring the gap from the majority of net surface charge, ganglioside-sialate at the edge of the hydration layer, but is separately motivated: The headgroups' dielectric profile is broadened by lipid protrusion \cite{Aniansson1978}, and by intrusion of ions and water molecules around the ester links to the tailgroups  \cite{1978Buldt,Lucas2012,Stern2003,Nymeyer2008}. Also, equivalent to dispersion pressure repelling two regions separated by medium with intermediate permittivity \cite{MahantyNinham1976}, dispersion will favour any average hydration layer structure with permittivity intermediate to the headgroups and the solution. For all these reasons, I take the hydration together with the rest of the headgroup interlayer to have the same permittivity and include its width in the calculation of this value: Modelling headgroup permittivity as a function of frequency requires estimates for the low-frequency relative permittivity, the refractive index in the visible spectrum, and the main electronic absorption frequency of ultraviolet light. The latter must be $\approx1.428\times10^{15}\text{ Hz}$, the peak UV absorption frequency of phospholipids \cite{Spector1996} that comprise two-thirds of the bilayer \cite{Breckenridge1972}. Headgroups’ low-frequency and optical permittivities must combine with hydrocarbon tails’ values to agree with overall bilayer measurements. Each band of the bilayer displays a capacitance proportional to the quotient of its permittivity by its thickness, and as these capacitances are in series the combining equation is $D_{B}/\epsilon_{B}=(2D_{H}/\epsilon_{H})+(2D_{T}/\epsilon_{T})$ where $D_{B}$ is the bilayer thickness and $D_{H}$ and $D_{T}$ are the thicknesses of the headgroups and tailgroups on each side, so that  $D_{B}=2D_{H}+2D_{T}$. Thus headgroup permittivity is $\epsilon_{H}=2D_{H}/[(D_{B}/\epsilon_{B})-(2D_{T}/\epsilon_{T})]$. In the zero-frequency regime the overall relative permittivity of lipid bilayer is 3.2, measured at 80 MHz \cite{Gramse2013} to avoid million-fold higher ionic polarization \cite{Maxwell1891,Gabriel1996}. This approximates the permittivity of the substance apparent to zero-frequency electromagnetic fluctuations limited in energy to around $k_{B}T$, evidently only sufficient there to polarize electron density and hydrogen bond screening. The permittivity of the hydrocarbon tails at the bilayer core is 2.0 \cite{Parsegian2005}. Molecular dynamics simulations \cite{Lucas2012,Stern2003,Nymeyer2008} agree with experimental measurements \cite{1978Buldt,1996Koenig} that ions and water molecules diffuse approximately 0.25 nm into the 1.25 nm tail moieties, around the ester links \cite{1978Buldt}, and that there is an $\approx0.75\text{ nm}$ \cite{1996Koenig} hydration layer associated with the $\approx1.0\text{ nm}$ headgroup layer \cite{1996Koenig}, making the relevant thicknesses $D_{H}\approx2.0\text{ nm}$ and $D_{T}\approx1.0\text{ nm}$ for a total bilayer thickness $D_{B}\approx6.0\text{ nm}$.
These parameters yield a low-frequency headgroup layer permittivity of $\varepsilon_{H}\approx4.6$. In the frequency range of visible light to ultraviolet the refractive indices are approximately 1.45 for the bilayer \cite{Ardhammar2002,2014Granqvist,2018Parkkila}  and 1.41 for the hydrocarbon tailgroups \cite{Israelachvili2011}, corresponding to permittivities of 2.1 for the overall bilayer and still 2.0 for the tailgroups. These estimates indicate the polar headgroup layer has an optical permittivity $\varepsilon_{H}\approx2.165$, hence refractive index $n_{H}\approx1.471$. For water, the permittivity is fitted in terms of $i\xi$ by \cite{Parsegian2005}:
\begin{equation}
\epsilon_{r}(i\xi)=1+\frac{d}{1+\xi\tau}+\sum\nolimits_{j}\frac{f_{j}}{\omega_{j}^{2}+\xi g_{j}+\xi^{2}}
\end{equation}
The microwave band is covered by parameters $d=74.8$, $1/\tau=9.95\times10^{10}\text{ rad s\textsuperscript{-1}}$ and higher angular frequencies $\xi$ by the other coefficients \cite{Parsegian2005}, absent $10^{15}$-$10^{16}\text{ rad s\textsuperscript{-1}}$ where water is transparent.

\begin{figure}
\includegraphics{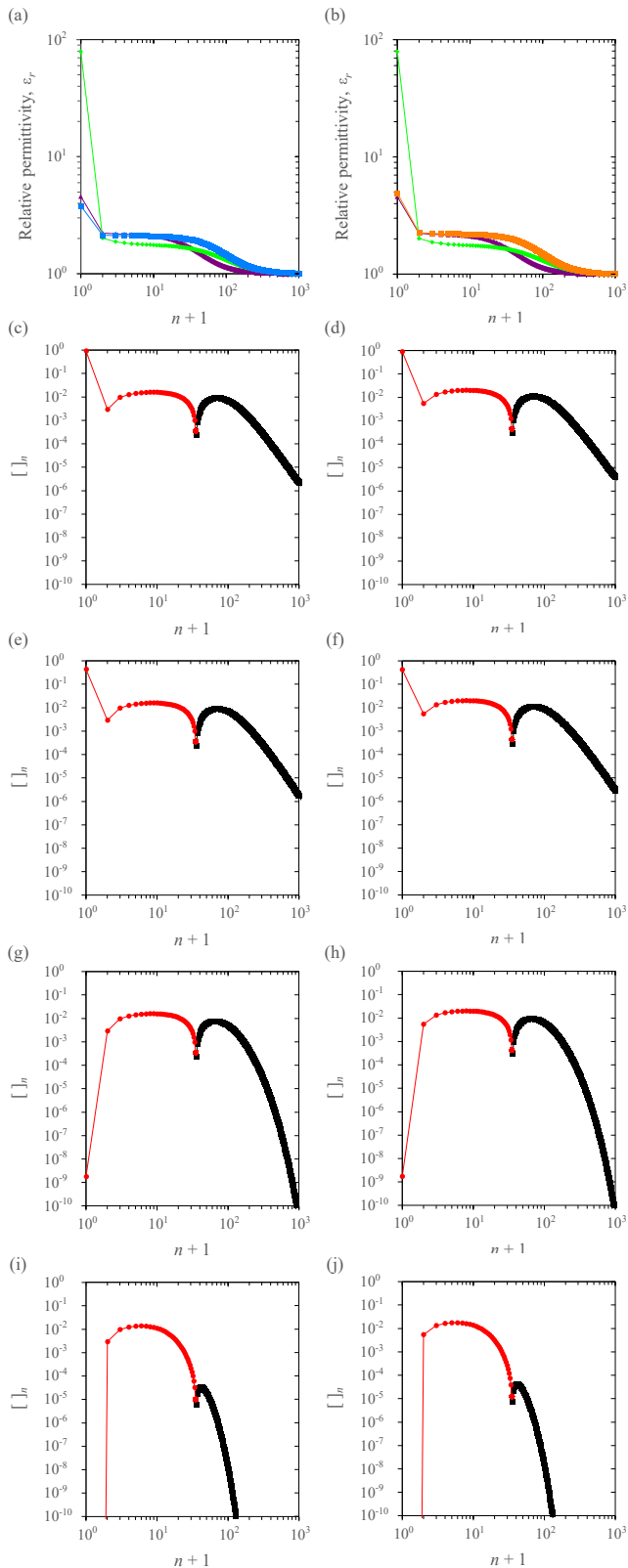}
\caption{
\label{fig:terms}Graphs illustrating the origin of the three regimes of dispersion pressure between glass and cell membrane. Left panels: Silicate. Right panels: Borosilicate. (a, b) Relative permittivities evaluated at $i\xi_{n}=2\pi i n k_{B}T/\hbar$. For $n \ge 35$ the saline medium (green diamonds) has intermediate permittiv-
}
\end{figure}

The dispersion pressure (Fig.~\ref{fig:graphs}) calculated from these permittivity functions  (Fig.~\ref{fig:terms}) repels the surfaces at intermediate separations: The strongly attractive $n=0$ term in Eq.~(\ref{eqn:DisP}) outweighs the net repulsion from the other terms, up to (0.5)0.8 nm for (boro)silicate, where its ion-screened decay begins to reveal this barrier. At wider separations the attractive terms $n\leq34$ desynchronize last, forming the loose adhesion regime. Conversely, symmetry guarantees entirely attractive dispersion pressure between pairs of bilayers \cite{MahantyNinham1976}.
%\newline

\textit{Net pressure.} The net pressure, the sum of the dispersion and electrostatic pressures, repels the surfaces at approach lengths $l$ between 2 nm and 18 nm, with peak barrier pressures of 6.3 kPa and 7.5 kPa for quartz and borosilicate at 2.63 nm and 2.55
 nm respectively, matching the experimental range of 5-10 kPa \cite{Yang2013,Ursell2011}. For high resistance, membrane must flatten  against the glass while sealing \cite{Ursell2011}, impossible without a pressure barrier. To reduce the tension perturbing ion channels \cite{Hamill1997,Ursell2011,Sachs2015} in a membrane patch from sealing its perimeter though, the model indicates lower glass surface charge would lower peak repulsion, by decreasing double-layer repulsion relative to calcium-bridge attraction: At glass surface charge densities of (-1.75) -1.42 mC/m\textsuperscript{2} Fig.~\ref{fig:graphs} shows joining pressure minima of (5.0) 3.4 kPa for (boro)silicate.

The pressure barrier is also seen by Quartz Crystal Microbalance with Dissipation monitoring (QCM-D) of 75-100~nm diameter unilamellar vesicles (prepared by sonication or extrusion \cite{2004Schonherr,2018Kurniawan}) on glass, as reversible loose adhesion for generally more natural mixed-lipid compositions \cite{2018Biswas,2006Richter}, modulated by calcium \cite{2003Richter} as expected. Loosely-adhered vesicles cannot be easily studied by Atomic Force Microscopy (AFM) because the cantilever tips apply enough pressure to push them into tight adhesion even in the gentlest scanning modes - The minimum AFM tip pressure, for 30 pN force and 30 nm tip radius \cite{2004Schonherr}, is $\approx10.6\text{ kPa}$. Similarly, in Langmuir-Blodgett deposition of Solid-supported Lipid Bilayer (SLB) \cite{1984Brian,1986McConnell,2018Kurniawan} initial edge-nucleation would usually ensure progressive mechanical propagation of adhesion from an interface.

Integrating the adhesive branch of the net pressure down to either the Debye length, 0.78 nm, or the unhydrated width of Na\textsuperscript{+}, 0.19 nm \cite{Israelachvili2011}, gives adhesion energies for silicate of 0.51 and 3.0 mJ/m\textsuperscript{2} respectively, agreeing with the experimental range of 0.4-4.0 mJ/m\textsuperscript{2} \cite{Ursell2011,Opsahl1994}. For borosilicate theory gives 0.49 and 2.6 mJ/m\textsuperscript{2}. The other attractive branch at far separations accounts for loose adhesion \cite{Almers1983}, integrating to weak association energies of 30 and 39 nJ/m\textsuperscript{2} for silicate and borosilicate. 
Neglecting ion screening and the finite velocity of light, near-field approximation can estimate tight adhesion \cite{Slavchov2014} but cannot predict any barrier or loose adhesion.

Even though the ion-correlated and electrodynamic pressures are both strongly attractive at close range, tight adhesion of cell membrane to glass does not happen instantaneously, with micropipet seals taking 2-30 s to exceed 1 G$\Omega$ resistance \cite{Hamill1981,Kolb2016}. This delay indicates that the variation within each laboratory's measurements of the adhesion energy, 0.5-4.0 mJ~m\textsuperscript{-2} \cite{Opsahl1994}, and 0.4-4.0 mJ~m\textsuperscript{-2} \cite{Ursell2011}, is from viscous surface hydrogen bonding slowing ion rearrangement, water diffusion, and hence gap constriction to such an extent that variation in settling time matters to the measurements. The slow lateral escape of excess water might also limit the mechanical propagation rate of adhesion fronts. At equilibrium tight adhesion, the combination of electrostatic and electrodynamic constraints on the hydration layer would frustrate ion traversal, accounting for the high electrical resistivity.

The process usually measured to gauge adhesion is the reverse, peeling, which requires only $\Delta p\lesssim2.5$ kPa on membranes across borosilicate pipets with inner radius $a\approx5.75$ $\mu$m \cite{Opsahl1994}.
Thus an upper estimate for the experimental detachment pressure from tight adhesion, assuming microscopic peeling force normal to the wall and equating integrals of line tension around the membrane circumferences facing upwards and outwards, is $\Delta p$ times the ratio of patch cross-section $\pi a^{2}$ to the incremental peel area $2\pi a/\sqrt{\Gamma}$, viz., $\sqrt{\Gamma}a\Delta p/2\approx9$ MPa. 
This agrees with the theoretical detachment pressure from borosilicate at the Na\textsuperscript{+} limit, 7.9 MPa, cf. Fig.~\ref{fig:graphs}. Detachment pressure exceeding the usual 3 MPa membrane rupture tension \cite{Needham1990} is consistent with pipet reuse requiring thorough cleaning \cite{Kolb2016}, and with how raising a tightly adhered pipet either excises the patch or uproots the cell, so peeling bilayer is evidently stabilized by the remaining surface adhesion.
%\newline

\textit{Steric renormalization.} Most stresses on a cell are borne by the cytoskeleton so there is usually slack in the membrane \cite{SachsSivaselvan2015} that allows it to undulate. If a membrane repels another surface at short range these undulations and the protrusions of its lipid molecules extend the distance profile of the pressure exerted. Various active cellular processes could excite lipid protrusions
\begin{minipage}[b]{\columnwidth}
%\flushing
\small\rmfamily
\noindent
\vspace{18pt}
\hrule
\vspace{18pt}
-ity to the glass (squares) and hydrated headgroups (purple triangles), so the corresponding terms in Eq. (8) contribute to pushing the surfaces apart. The lower panels show the variation with approach length \textit{l} of the first thousand terms in square brackets of Eq. (8), without the prefactor of $k_{B}T/8\pi l^{3}$, to aid comparison, but including the overall minus sign, so that negative terms (red circles) signify attractive contributions to the overall pressure while the positive terms (black squares) contribute to pushing the surfaces apart. (c, d) $l=0.1$ nm, (e, f) $l=1$ nm, (g, h) $l=10$ nm, (i, j) $l=100$ nm. The ion-screened $n=0$ term contributes a strongly attractive pressure exceeding all the other terms combined at $l=0.1$ nm, but by $l=1$ nm has decayed sufficiently for the sum of the negative terms $n\le34$ to no longer outweigh the sum of the positive terms $n\ge35$. As the $n=0$ term continues to decay, the dispersion barrier is progressively revealed. However, with further separation, relativistic desynchronization weakens higher-order terms first until, soon after $l=10$ nm, the pressure switches sign again, because the attractive terms $n\le34$ are yet to desynchronize.
\end{minipage}
and bilayer 
undulations too, but as far as current estimations are concerned the excitations are entirely thermal. The pressure
on a wall from thermal undulations is   $0.21k_{B}^{2}T^{2}/CD^{3}$ \cite{Helfrich1978}
for peak depth \textit{D} and curvature modulus $C\approx20\text{zJ}$  \cite{Israelachvili2011} (lower at the melting transition \cite{Dimova2014}). Pressure 
from lipid protrusion, previously determined for pairs of bilayers \cite{Aniansson1978}, whose hydration is taken to repel them conservatively at short range \cite{Israelachvili2011}, I derive here for a single bilayer next to a wall in the same way. The derivations use the potential distribution theorem that $\langle w(z) \rangle\approx-kT \ln(\int\text{e}^{-w(z,x)/kT}dx/\int dx)$ and renormalize to accessible states by declaring an arbitrary practical limit of the state space to be $L$, a few multiples of the protrusion decay length $\lambda = kT/\alpha$, that differentiates out: The interaction free energy per site at wall separation $D$ given the protrusion potential $\alpha z$ is $\langle w(D) \rangle \approx -kT \ln(\int_{0}^{D} \text{e}^{-\alpha h/kT}dh/\int_{0}^{L} dx) = -kT\ln[\lambda(1-\text{e}^{-D/\lambda})/L]$ and hence the protrusion pressure on a wall is $P = -\partial\Gamma\langle w(D)\rangle/\partial D = (\Gamma\alpha\text{e}^{-D/\lambda})/(1-\text{e}^{-D/\lambda})$ where $\alpha\approx2.5\times10^{-11}\text{ J m\textsuperscript{-1}}$  \cite{Israelachvili2011} and $\lambda\approx0.165\text{ nm}$  \cite{Aniansson1978}. The sum of these two steric pressures renormalizes the barrier by extending each \textit{l} by the corresponding \textit{D}. Renormalized numerically, the pressure barriers become approximately inversely proportional to the third power of the mean gap \textit{g}, matching recent experimental measurements of non-contact cell membrane compression by glass \cite{Clarke2016}, and specifically fitted by Table~\ref{tab:table2}'s parameters for
\begin{equation}
P(g)=\frac{1}{6\pi}\frac{h-fg}{v+(g-d)^{3}}.
\end{equation}
\begin{table}%[]
\caption{\label{tab:table2}Renormalized pressure barrier fits and limits.}
\begin{ruledtabular}
\begin{tabular}{lccccll}
 & \textit{f} & \textit{v} & \textit{h} & \textit{d} & $g_{min}$ & $g_{max}$ \\
 &(fN)&\multicolumn{1}{l}{(nm\textsuperscript{3})}&(zJ)&(nm)&(nm)&(nm)\\
\colrule
Borosilicate & 74.4 & 58.6 & 13.1 & 2.39 & 5.39 & 123\\
Silicate & 59.9 & 63.3 & 11.9 & 2.50 & 5.62 & 135\\
\end{tabular}
\end{ruledtabular}
\end{table}

\textit{Conclusions.} The most striking aspect of the model is the theoretical unification of the repulsion with loose and tight adhesion. By rendering four associated measurements, the adhesion energies per unit area, detachment pressure and peak repulsion, the model shows gangliosides stabilize outer bilayer hydration and that tight adhesion slowly progresses from salt bridges to electrostatic complementarity. Of further significance to electrophysiology is the lower peak repulsion calculated for silicate glass, an unrecognized advantage for reducing patch tension and hence ion channel perturbation \cite{Hamill1997,Sachs2015} that the model predicts would improve with lower glass surface charge density.

Quantitatively resolving phenomena fundamental to biology, the theory predicts cells forming biofilms on silicates must first flatten to maximize force from loose adhesion so active protrusions can push through the barrier to tight adhesion. Conversely, in relation to enveloped virions’ passive interactions, the model rationalizes the hitherto empirical use of glass vials for storing such viruses \cite{2017Newcomb} and vaccines based on them \cite{2016White}: Surface association by loose adhesion of envelope to glass may require displacement by vial inversion, but nonspecific spike interactions are too transient, weak and offset to pull envelopes into tight adhesion. Furthermore, the estimate of ion correlation pressure indicates that to enhance storage by maintaining the highest possible pressure barrier to tight adhesion, buffer formulations should not include multivalent cations like Ca\textsuperscript{2+}. More generally, the theory shows accessible measurements of the asymmetric membrane-saline-glass system can accurately refine tractable models of lipid bilayer that are directly transferable to cell membrane interactions like synaptogenesis \cite{Betz2011}. 
%\appendix*
%\section{}
\begin{acknowledgments}
I thank Salome Antolin, Pavel Novak, and David Klenerman for insightful discussions. This research was supported by the University of Cambridge, Christ’s College, the BBSRC (BB/L006227/1), EPSRC (EP/H01098X/1), and NPL.
\end{acknowledgments}

% \bibliography{basename of .bib file}

\bibliography{TCMIG}

\begin{thebibliography}{64}%
\makeatletter
\providecommand \@ifxundefined [1]{%
 \@ifx{#1\undefined}
}%
\providecommand \@ifnum [1]{%
 \ifnum #1\expandafter \@firstoftwo
 \else \expandafter \@secondoftwo
 \fi
}%
\providecommand \@ifx [1]{%
 \ifx #1\expandafter \@firstoftwo
 \else \expandafter \@secondoftwo
 \fi
}%
\providecommand \natexlab [1]{#1}%
\providecommand \enquote  [1]{``#1''}%
\providecommand \bibnamefont  [1]{#1}%
\providecommand \bibfnamefont [1]{#1}%
\providecommand \citenamefont [1]{#1}%
\providecommand \href@noop [0]{\@secondoftwo}%
\providecommand \href [0]{\begingroup \@sanitize@url \@href}%
\providecommand \@href[1]{\@@startlink{#1}\@@href}%
\providecommand \@@href[1]{\endgroup#1\@@endlink}%
\providecommand \@sanitize@url [0]{\catcode `\\12\catcode `\$12\catcode
  `\&12\catcode `\#12\catcode `\^12\catcode `\_12\catcode `\%12\relax}%
\providecommand \@@startlink[1]{}%
\providecommand \@@endlink[0]{}%
\providecommand \url  [0]{\begingroup\@sanitize@url \@url }%
\providecommand \@url [1]{\endgroup\@href {#1}{\urlprefix }}%
\providecommand \urlprefix  [0]{URL }%
\providecommand \Eprint [0]{\href }%
\providecommand \doibase [0]{https://doi.org/}%
\providecommand \selectlanguage [0]{\@gobble}%
\providecommand \bibinfo  [0]{\@secondoftwo}%
\providecommand \bibfield  [0]{\@secondoftwo}%
\providecommand \translation [1]{[#1]}%
\providecommand \BibitemOpen [0]{}%
\providecommand \bibitemStop [0]{}%
\providecommand \bibitemNoStop [0]{.\EOS\space}%
\providecommand \EOS [0]{\spacefactor3000\relax}%
\providecommand \BibitemShut  [1]{\csname bibitem#1\endcsname}%
\let\auto@bib@innerbib\@empty
%</preamble>
\bibitem [{\citenamefont {Hamill}\ \emph {et~al.}(1981)\citenamefont {Hamill},
  \citenamefont {Marty}, \citenamefont {Neher}, \citenamefont {Sakmann},\ and\
  \citenamefont {Sigworth}}]{Hamill1981}%
  \BibitemOpen
  \bibfield  {author} {\bibinfo {author} {\bibfnamefont {O.~P.}\ \bibnamefont
  {Hamill}}, \bibinfo {author} {\bibfnamefont {A.}~\bibnamefont {Marty}},
  \bibinfo {author} {\bibfnamefont {E.}~\bibnamefont {Neher}}, \bibinfo
  {author} {\bibfnamefont {B.}~\bibnamefont {Sakmann}},\ and\ \bibinfo {author}
  {\bibfnamefont {F.~J.}\ \bibnamefont {Sigworth}},\ }\bibfield  {title}
  {\bibinfo {title} {Improved patch-clamp techniques for high-resolution
  current recording from cells and cell-free membrane patches},\ }\href
  {https://doi.org/10.1007/BF00656997} {\bibfield  {journal} {\bibinfo
  {journal} {Pfl{\"u}gers Archiv}\ }\textbf {\bibinfo {volume} {391}},\
  \bibinfo {pages} {85} (\bibinfo {year} {1981})},\ \Eprint
  {https://arxiv.org/abs/10.1007/BF00656997} {10.1007/BF00656997} \BibitemShut
  {NoStop}%
\bibitem [{\citenamefont {Almers}\ \emph {et~al.}(1983)\citenamefont {Almers},
  \citenamefont {Stanfield},\ and\ \citenamefont {St{\"u}hmer}}]{Almers1983}%
  \BibitemOpen
  \bibfield  {author} {\bibinfo {author} {\bibfnamefont {W.}~\bibnamefont
  {Almers}}, \bibinfo {author} {\bibfnamefont {P.}~\bibnamefont {Stanfield}},\
  and\ \bibinfo {author} {\bibfnamefont {W.}~\bibnamefont {St{\"u}hmer}},\
  }\bibfield  {title} {\bibinfo {title} {Lateral distribution of sodium and
  potassium channels in frog skeletal muscle: measurements with a patch-clamp
  technique.},\ }\href {https://doi.org/10.1113/jphysiol.1983.sp014580}
  {\bibfield  {journal} {\bibinfo  {journal} {J. Physiol.}\ }\textbf {\bibinfo
  {volume} {336}},\ \bibinfo {pages} {261} (\bibinfo {year} {1983})},\ \Eprint
  {https://arxiv.org/abs/10.1113/jphysiol.1983.sp014580}
  {10.1113/jphysiol.1983.sp014580} \BibitemShut {NoStop}%
\bibitem [{\citenamefont {Opsahl}\ and\ \citenamefont
  {Webb}(1994)}]{Opsahl1994}%
  \BibitemOpen
  \bibfield  {author} {\bibinfo {author} {\bibfnamefont {L.}~\bibnamefont
  {Opsahl}}\ and\ \bibinfo {author} {\bibfnamefont {W.}~\bibnamefont {Webb}},\
  }\bibfield  {title} {\bibinfo {title} {Lipid-glass adhesion in giga-sealed
  patch-clamped membranes},\ }\href
  {https://doi.org/10.1016/S0006-3495(94)80752-0} {\bibfield  {journal}
  {\bibinfo  {journal} {Biophys. J.}\ }\textbf {\bibinfo {volume} {66}},\
  \bibinfo {pages} {75 } (\bibinfo {year} {1994})},\ \Eprint
  {https://arxiv.org/abs/10.1016/S0006-3495(94)80752-0}
  {10.1016/S0006-3495(94)80752-0} \BibitemShut {NoStop}%
\bibitem [{\citenamefont {Ursell}\ \emph {et~al.}(2011)\citenamefont {Ursell},
  \citenamefont {Agrawal},\ and\ \citenamefont {Phillips}}]{Ursell2011}%
  \BibitemOpen
  \bibfield  {author} {\bibinfo {author} {\bibfnamefont {T.}~\bibnamefont
  {Ursell}}, \bibinfo {author} {\bibfnamefont {A.}~\bibnamefont {Agrawal}},\
  and\ \bibinfo {author} {\bibfnamefont {R.}~\bibnamefont {Phillips}},\
  }\bibfield  {title} {\bibinfo {title} {Lipid bilayer mechanics in a pipette
  with glass-bilayer adhesion},\ }\href
  {https://doi.org/10.1016/j.bpj.2011.08.057} {\bibfield  {journal} {\bibinfo
  {journal} {Biophys. J.}\ }\textbf {\bibinfo {volume} {101}},\ \bibinfo
  {pages} {1913 } (\bibinfo {year} {2011})},\ \Eprint
  {https://arxiv.org/abs/10.1016/j.bpj.2011.08.057} {10.1016/j.bpj.2011.08.057}
  \BibitemShut {NoStop}%
\bibitem [{\citenamefont {Hansma}\ \emph {et~al.}(1989)\citenamefont {Hansma},
  \citenamefont {Drake}, \citenamefont {Marti}, \citenamefont {Gould},\ and\
  \citenamefont {Prater}}]{Hansma1989}%
  \BibitemOpen
  \bibfield  {author} {\bibinfo {author} {\bibfnamefont {P.}~\bibnamefont
  {Hansma}}, \bibinfo {author} {\bibfnamefont {B.}~\bibnamefont {Drake}},
  \bibinfo {author} {\bibfnamefont {O.}~\bibnamefont {Marti}}, \bibinfo
  {author} {\bibfnamefont {S.}~\bibnamefont {Gould}},\ and\ \bibinfo {author}
  {\bibfnamefont {C.}~\bibnamefont {Prater}},\ }\bibfield  {title} {\bibinfo
  {title} {The scanning ion-conductance microscope},\ }\href
  {https://doi.org/10.1126/science.2464851} {\bibfield  {journal} {\bibinfo
  {journal} {Science}\ }\textbf {\bibinfo {volume} {243}},\ \bibinfo {pages}
  {641} (\bibinfo {year} {1989})},\ \Eprint
  {https://arxiv.org/abs/10.1126/science.2464851} {10.1126/science.2464851}
  \BibitemShut {NoStop}%
\bibitem [{\citenamefont {Clarke}\ \emph {et~al.}(2016)\citenamefont {Clarke},
  \citenamefont {Novak}, \citenamefont {Zhukov}, \citenamefont {Tyler},
  \citenamefont {Cano-Jaimez}, \citenamefont {Drews}, \citenamefont {Richards},
  \citenamefont {Volynski}, \citenamefont {Bishop},\ and\ \citenamefont
  {Klenerman}}]{Clarke2016}%
  \BibitemOpen
  \bibfield  {author} {\bibinfo {author} {\bibfnamefont {R.~W.}\ \bibnamefont
  {Clarke}}, \bibinfo {author} {\bibfnamefont {P.}~\bibnamefont {Novak}},
  \bibinfo {author} {\bibfnamefont {A.}~\bibnamefont {Zhukov}}, \bibinfo
  {author} {\bibfnamefont {E.~J.}\ \bibnamefont {Tyler}}, \bibinfo {author}
  {\bibfnamefont {M.}~\bibnamefont {Cano-Jaimez}}, \bibinfo {author}
  {\bibfnamefont {A.}~\bibnamefont {Drews}}, \bibinfo {author} {\bibfnamefont
  {O.}~\bibnamefont {Richards}}, \bibinfo {author} {\bibfnamefont
  {K.}~\bibnamefont {Volynski}}, \bibinfo {author} {\bibfnamefont
  {C.}~\bibnamefont {Bishop}},\ and\ \bibinfo {author} {\bibfnamefont
  {D.}~\bibnamefont {Klenerman}},\ }\bibfield  {title} {\bibinfo {title} {Low
  stress ion conductance microscopy of sub-cellular stiffness},\ }\href
  {https://doi.org/10.1039/C6SM01106C} {\bibfield  {journal} {\bibinfo
  {journal} {Soft Matter}\ }\textbf {\bibinfo {volume} {12}},\ \bibinfo {pages}
  {7953} (\bibinfo {year} {2016})},\ \Eprint
  {https://arxiv.org/abs/10.1039/C6SM01106C} {10.1039/C6SM01106C} \BibitemShut
  {NoStop}%
\bibitem [{\citenamefont {{Yang}}\ \emph {et~al.}(2013)\citenamefont {{Yang}},
  \citenamefont {{Lai}}, \citenamefont {{Xi}},\ and\ \citenamefont
  {{Yang}}}]{Yang2013}%
  \BibitemOpen
  \bibfield  {author} {\bibinfo {author} {\bibfnamefont {R.}~\bibnamefont
  {{Yang}}}, \bibinfo {author} {\bibfnamefont {K.~W.~C.}\ \bibnamefont
  {{Lai}}}, \bibinfo {author} {\bibfnamefont {N.}~\bibnamefont {{Xi}}},\ and\
  \bibinfo {author} {\bibfnamefont {J.}~\bibnamefont {{Yang}}},\ }\bibfield
  {title} {\bibinfo {title} {Development of automated patch clamp system for
  electrophysiology},\ }in\ \href {https://doi.org/10.1109/ROBIO.2013.6739793}
  {\emph {\bibinfo {booktitle} {2013 IEEE International Conference on Robotics
  and Biomimetics (ROBIO)}}}\ (\bibinfo  {publisher} {IEEE},\ \bibinfo {year}
  {2013})\ pp.\ \bibinfo {pages} {2185--2190},\ \Eprint
  {https://arxiv.org/abs/10.1109/ROBIO.2013.6739793}
  {10.1109/ROBIO.2013.6739793} \BibitemShut {NoStop}%
\bibitem [{\citenamefont {Hamill}\ and\ \citenamefont
  {McBride~Jr}(1997)}]{Hamill1997}%
  \BibitemOpen
  \bibfield  {author} {\bibinfo {author} {\bibfnamefont {O.~P.}\ \bibnamefont
  {Hamill}}\ and\ \bibinfo {author} {\bibfnamefont {D.~W.}\ \bibnamefont
  {McBride~Jr}},\ }\bibfield  {title} {\bibinfo {title} {Induced membrane
  hypo/hyper-mechanosensitivity: a limitation of patch-clamp recording},\
  }\href {https://doi.org/10.1146/annurev.physiol.59.1.621} {\bibfield
  {journal} {\bibinfo  {journal} {Annu. Rev. Physiol.}\ }\textbf {\bibinfo
  {volume} {59}},\ \bibinfo {pages} {621} (\bibinfo {year} {1997})},\ \Eprint
  {https://arxiv.org/abs/10.1146/annurev.physiol.59.1.621}
  {10.1146/annurev.physiol.59.1.621} \BibitemShut {NoStop}%
\bibitem [{\citenamefont {Sachs}(2015)}]{Sachs2015}%
  \BibitemOpen
  \bibfield  {author} {\bibinfo {author} {\bibfnamefont {F.}~\bibnamefont
  {Sachs}},\ }\bibfield  {title} {\bibinfo {title} {Mechanical transduction by
  ion channels: A cautionary tale},\ }\href
  {https://doi.org/10.5316/wjn.v5.i3.74} {\bibfield  {journal} {\bibinfo
  {journal} {World Journal of Neurology}\ }\textbf {\bibinfo {volume} {5}},\
  \bibinfo {pages} {74} (\bibinfo {year} {2015})},\ \Eprint
  {https://arxiv.org/abs/10.5316/wjn.v5.i3.74} {10.5316/wjn.v5.i3.74}
  \BibitemShut {NoStop}%
\bibitem [{\citenamefont {Hamill}(2014)}]{Hamill2014}%
  \BibitemOpen
  \bibfield  {author} {\bibinfo {author} {\bibfnamefont {O.~P.}\ \bibnamefont
  {Hamill}},\ }\bibinfo {title} {Patch-clamp technique},\ in\ \href
  {https://doi.org/10.1002/9780470015902.a0003382.pub2} {\emph {\bibinfo
  {booktitle} {eLS}}}\ (\bibinfo  {publisher} {American Cancer Society},\
  \bibinfo {year} {2014})\ Chap.\ \bibinfo {chapter} {Abstr.}, pp.\ \bibinfo
  {pages} {1--10},\ \Eprint
  {https://arxiv.org/abs/10.1002/9780470015902.a0003382.pub2}
  {10.1002/9780470015902.a0003382.pub2} \BibitemShut {NoStop}%
\bibitem [{\citenamefont {Bae}\ \emph {et~al.}(2011)\citenamefont {Bae},
  \citenamefont {Markin}, \citenamefont {Suchyna},\ and\ \citenamefont
  {Sachs}}]{Bae2011}%
  \BibitemOpen
  \bibfield  {author} {\bibinfo {author} {\bibfnamefont {C.}~\bibnamefont
  {Bae}}, \bibinfo {author} {\bibfnamefont {V.}~\bibnamefont {Markin}},
  \bibinfo {author} {\bibfnamefont {T.}~\bibnamefont {Suchyna}},\ and\ \bibinfo
  {author} {\bibfnamefont {F.}~\bibnamefont {Sachs}},\ }\bibfield  {title}
  {\bibinfo {title} {Modeling ion channels in the gigaseal},\ }\href
  {https://doi.org/https://dx.doi.org/10.1016/j.bpj.2011.11.002} {\bibfield
  {journal} {\bibinfo  {journal} {Biophys. J.}\ }\textbf {\bibinfo {volume}
  {101}},\ \bibinfo {pages} {2645 } (\bibinfo {year} {2011})},\ \Eprint
  {https://arxiv.org/abs/10.1016/j.bpj.2011.11.002} {10.1016/j.bpj.2011.11.002}
  \BibitemShut {NoStop}%
\bibitem [{\citenamefont {Novak}\ \emph {et~al.}(2013)\citenamefont {Novak},
  \citenamefont {Gorelik}, \citenamefont {Vivekananda}, \citenamefont
  {Shevchuk}, \citenamefont {Ermolyuk}, \citenamefont {Bailey}, \citenamefont
  {Bushby}, \citenamefont {Moss}, \citenamefont {Rusakov}, \citenamefont
  {Klenerman}, \citenamefont {Kullmann}, \citenamefont {Volynski},\ and\
  \citenamefont {Korchev}}]{Novak2013}%
  \BibitemOpen
  \bibfield  {author} {\bibinfo {author} {\bibfnamefont {P.}~\bibnamefont
  {Novak}}, \bibinfo {author} {\bibfnamefont {J.}~\bibnamefont {Gorelik}},
  \bibinfo {author} {\bibfnamefont {U.}~\bibnamefont {Vivekananda}}, \bibinfo
  {author} {\bibfnamefont {A.~I.}\ \bibnamefont {Shevchuk}}, \bibinfo {author}
  {\bibfnamefont {Y.~S.}\ \bibnamefont {Ermolyuk}}, \bibinfo {author}
  {\bibfnamefont {R.~J.}\ \bibnamefont {Bailey}}, \bibinfo {author}
  {\bibfnamefont {A.~J.}\ \bibnamefont {Bushby}}, \bibinfo {author}
  {\bibfnamefont {G.~W.}\ \bibnamefont {Moss}}, \bibinfo {author}
  {\bibfnamefont {D.~A.}\ \bibnamefont {Rusakov}}, \bibinfo {author}
  {\bibfnamefont {D.}~\bibnamefont {Klenerman}}, \bibinfo {author}
  {\bibfnamefont {D.~M.}\ \bibnamefont {Kullmann}}, \bibinfo {author}
  {\bibfnamefont {K.~E.}\ \bibnamefont {Volynski}},\ and\ \bibinfo {author}
  {\bibfnamefont {Y.~E.}\ \bibnamefont {Korchev}},\ }\bibfield  {title}
  {\bibinfo {title} {Nanoscale-targeted patch-clamp recordings of functional
  presynaptic ion channels},\ }\href
  {https://doi.org/https://doi.org/10.1016/j.neuron.2013.07.012} {\bibfield
  {journal} {\bibinfo  {journal} {Neuron}\ }\textbf {\bibinfo {volume} {79}},\
  \bibinfo {pages} {1067 } (\bibinfo {year} {2013})},\ \Eprint
  {https://arxiv.org/abs/10.1016/j.neuron.2013.07.012}
  {10.1016/j.neuron.2013.07.012} \BibitemShut {NoStop}%
\bibitem [{\citenamefont {Ogden}\ and\ \citenamefont
  {Stanfield}(1994)}]{Ogden1994}%
  \BibitemOpen
  \bibfield  {author} {\bibinfo {author} {\bibfnamefont {D.}~\bibnamefont
  {Ogden}}\ and\ \bibinfo {author} {\bibfnamefont {P.}~\bibnamefont
  {Stanfield}},\ }\href@noop {} {\emph {\bibinfo {title} {Microelectrode
  techniques}}}\ (\bibinfo  {publisher} {The Company of Biologists Ltd},\
  \bibinfo {year} {1994})\ Chap.\ \bibinfo {chapter} {Patch clamp techniques
  for single channel and whole-cell recording}, pp.\ \bibinfo {pages}
  {53--78}\BibitemShut {NoStop}%
\bibitem [{\citenamefont {J{\"o}nsson}\ \emph {et~al.}(2012)\citenamefont
  {J{\"o}nsson}, \citenamefont {McColl}, \citenamefont {Clarke}, \citenamefont
  {Ostanin}, \citenamefont {J{\"o}nsson},\ and\ \citenamefont
  {Klenerman}}]{Jonsson2012}%
  \BibitemOpen
  \bibfield  {author} {\bibinfo {author} {\bibfnamefont {P.}~\bibnamefont
  {J{\"o}nsson}}, \bibinfo {author} {\bibfnamefont {J.}~\bibnamefont {McColl}},
  \bibinfo {author} {\bibfnamefont {R.~W.}\ \bibnamefont {Clarke}}, \bibinfo
  {author} {\bibfnamefont {V.~P.}\ \bibnamefont {Ostanin}}, \bibinfo {author}
  {\bibfnamefont {B.}~\bibnamefont {J{\"o}nsson}},\ and\ \bibinfo {author}
  {\bibfnamefont {D.}~\bibnamefont {Klenerman}},\ }\bibfield  {title} {\bibinfo
  {title} {Hydrodynamic trapping of molecules in lipid bilayers},\ }\href
  {https://doi.org/10.1073/pnas.1202858109} {\bibfield  {journal} {\bibinfo
  {journal} {Proc. Nat. Acad. Sci. U.S.A.}\ }\textbf {\bibinfo {volume}
  {109}},\ \bibinfo {pages} {10328} (\bibinfo {year} {2012})},\ \Eprint
  {https://arxiv.org/abs/10.1073/pnas.1202858109} {10.1073/pnas.1202858109}
  \BibitemShut {NoStop}%
\bibitem [{\citenamefont {Suchyna}\ \emph {et~al.}(2009)\citenamefont
  {Suchyna}, \citenamefont {Markin},\ and\ \citenamefont
  {Sachs}}]{Suchyna2009}%
  \BibitemOpen
  \bibfield  {author} {\bibinfo {author} {\bibfnamefont {T.~M.}\ \bibnamefont
  {Suchyna}}, \bibinfo {author} {\bibfnamefont {V.~S.}\ \bibnamefont
  {Markin}},\ and\ \bibinfo {author} {\bibfnamefont {F.}~\bibnamefont
  {Sachs}},\ }\bibfield  {title} {\bibinfo {title} {Biophysics and structure of
  the patch and the gigaseal},\ }\href
  {https://doi.org/10.1016/j.bpj.2009.05.018} {\bibfield  {journal} {\bibinfo
  {journal} {Biophys. J.}\ }\textbf {\bibinfo {volume} {97}},\ \bibinfo {pages}
  {738 } (\bibinfo {year} {2009})},\ \Eprint
  {https://arxiv.org/abs/10.1016/j.bpj.2009.05.018} {10.1016/j.bpj.2009.05.018}
  \BibitemShut {NoStop}%
\bibitem [{\citenamefont {Rheinlaender}\ and\ \citenamefont
  {Sch{\"a}ffer}(2013)}]{Rheinlaender2013}%
  \BibitemOpen
  \bibfield  {author} {\bibinfo {author} {\bibfnamefont {J.}~\bibnamefont
  {Rheinlaender}}\ and\ \bibinfo {author} {\bibfnamefont {T.~E.}\ \bibnamefont
  {Sch{\"a}ffer}},\ }\bibfield  {title} {\bibinfo {title} {Mapping the
  mechanical stiffness of live cells with the scanning ion conductance
  microscope},\ }\href {https://doi.org/10.1039/C2SM27412D} {\bibfield
  {journal} {\bibinfo  {journal} {Soft Matter}\ }\textbf {\bibinfo {volume}
  {9}},\ \bibinfo {pages} {3230} (\bibinfo {year} {2013})},\ \Eprint
  {https://arxiv.org/abs/10.1039/C2SM27412D} {10.1039/C2SM27412D} \BibitemShut
  {NoStop}%
\bibitem [{\citenamefont {Sachs}\ and\ \citenamefont
  {Sivaselvan}(2015)}]{SachsSivaselvan2015}%
  \BibitemOpen
  \bibfield  {author} {\bibinfo {author} {\bibfnamefont {F.}~\bibnamefont
  {Sachs}}\ and\ \bibinfo {author} {\bibfnamefont {M.~V.}\ \bibnamefont
  {Sivaselvan}},\ }\bibfield  {title} {\bibinfo {title} {Cell volume control in
  three dimensions: Water movement without solute movement},\ }\href
  {https://doi.org/10.1085/jgp.201411297} {\bibfield  {journal} {\bibinfo
  {journal} {J. Gen. Physiol.}\ }\textbf {\bibinfo {volume} {145}},\ \bibinfo
  {pages} {373} (\bibinfo {year} {2015})},\ \Eprint
  {https://arxiv.org/abs/10.1085/jgp.201411297} {10.1085/jgp.201411297}
  \BibitemShut {NoStop}%
\bibitem [{\citenamefont {Clarke}\ \emph {et~al.}(2013)\citenamefont {Clarke},
  \citenamefont {Zhukov}, \citenamefont {Richards}, \citenamefont {Johnson},
  \citenamefont {Ostanin},\ and\ \citenamefont {Klenerman}}]{Clarke2013}%
  \BibitemOpen
  \bibfield  {author} {\bibinfo {author} {\bibfnamefont {R.~W.}\ \bibnamefont
  {Clarke}}, \bibinfo {author} {\bibfnamefont {A.}~\bibnamefont {Zhukov}},
  \bibinfo {author} {\bibfnamefont {O.}~\bibnamefont {Richards}}, \bibinfo
  {author} {\bibfnamefont {N.}~\bibnamefont {Johnson}}, \bibinfo {author}
  {\bibfnamefont {V.}~\bibnamefont {Ostanin}},\ and\ \bibinfo {author}
  {\bibfnamefont {D.}~\bibnamefont {Klenerman}},\ }\bibfield  {title} {\bibinfo
  {title} {Pipette–surface interaction: Current enhancement and intrinsic
  force},\ }\href {https://doi.org/10.1021/ja3094586} {\bibfield  {journal}
  {\bibinfo  {journal} {J. Am. Chem. Soc.}\ }\textbf {\bibinfo {volume}
  {135}},\ \bibinfo {pages} {SI} (\bibinfo {year} {2013})},\ \Eprint
  {https://arxiv.org/abs/10.1021/ja3094586} {10.1021/ja3094586} \BibitemShut
  {NoStop}%
\bibitem [{\citenamefont {Mahanty}\ and\ \citenamefont
  {Ninham}(1976)}]{MahantyNinham1976}%
  \BibitemOpen
  \bibfield  {author} {\bibinfo {author} {\bibfnamefont {J.}~\bibnamefont
  {Mahanty}}\ and\ \bibinfo {author} {\bibfnamefont {B.~W.}\ \bibnamefont
  {Ninham}},\ }\href@noop {} {\emph {\bibinfo {title} {Dispersion forces}}}\
  (\bibinfo  {publisher} {Academic Press},\ \bibinfo {address} {London},\
  \bibinfo {year} {1976})\BibitemShut {NoStop}%
\bibitem [{\citenamefont {Parsegian}(2005)}]{Parsegian2005}%
  \BibitemOpen
  \bibfield  {author} {\bibinfo {author} {\bibfnamefont {V.~A.}\ \bibnamefont
  {Parsegian}},\ }\href@noop {} {\emph {\bibinfo {title} {Van der Waals
  forces}}}\ (\bibinfo  {publisher} {Cambridge University Press},\ \bibinfo
  {address} {New York},\ \bibinfo {year} {2005})\BibitemShut {NoStop}%
\bibitem [{\citenamefont {Aniansson}(1978)}]{Aniansson1978}%
  \BibitemOpen
  \bibfield  {author} {\bibinfo {author} {\bibfnamefont {G.~E.}\ \bibnamefont
  {Aniansson}},\ }\bibfield  {title} {\bibinfo {title} {Dynamics and structure
  of micelles and other amphiphile structures},\ }\href
  {https://doi.org/10.1021/j100515a011} {\bibfield  {journal} {\bibinfo
  {journal} {J. Phys. Chem.}\ }\textbf {\bibinfo {volume} {82}},\ \bibinfo
  {pages} {2805} (\bibinfo {year} {1978})},\ \Eprint
  {https://arxiv.org/abs/10.1021/j100515a011} {10.1021/j100515a011}
  \BibitemShut {NoStop}%
\bibitem [{\citenamefont {Helfrich}(1978)}]{Helfrich1978}%
  \BibitemOpen
  \bibfield  {author} {\bibinfo {author} {\bibfnamefont {W.}~\bibnamefont
  {Helfrich}},\ }\bibfield  {title} {\bibinfo {title} {Steric interaction of
  fluid membranes in multilayer systems},\ }\href
  {https://doi.org/10.1515/zna-1978-0308} {\bibfield  {journal} {\bibinfo
  {journal} {Zeitschrift für Naturforschung A}\ }\textbf {\bibinfo {volume}
  {33}},\ \bibinfo {pages} {305} (\bibinfo {year} {1978})},\ \Eprint
  {https://arxiv.org/abs/10.1515/zna-1978-0308} {10.1515/zna-1978-0308}
  \BibitemShut {NoStop}%
\bibitem [{\citenamefont {Jednačak}\ \emph {et~al.}(1974)\citenamefont
  {Jednačak}, \citenamefont {Pravdić},\ and\ \citenamefont
  {Haller}}]{Jednacak1974}%
  \BibitemOpen
  \bibfield  {author} {\bibinfo {author} {\bibfnamefont {J.}~\bibnamefont
  {Jednačak}}, \bibinfo {author} {\bibfnamefont {V.}~\bibnamefont
  {Pravdić}},\ and\ \bibinfo {author} {\bibfnamefont {W.}~\bibnamefont
  {Haller}},\ }\bibfield  {title} {\bibinfo {title} {The electrokinetic
  potential of glasses in aqueous electrolyte solutions},\ }\href
  {https://doi.org/10.1016/0021-9797(74)90293-8} {\bibfield  {journal}
  {\bibinfo  {journal} {J. Colloid Interface Sci.}\ }\textbf {\bibinfo {volume}
  {49}},\ \bibinfo {pages} {16} (\bibinfo {year} {1974})},\ \Eprint
  {https://arxiv.org/abs/10.1016/0021-9797(74)90293-8}
  {10.1016/0021-9797(74)90293-8} \BibitemShut {NoStop}%
\bibitem [{\citenamefont {Breckenridge}\ \emph {et~al.}(1972)\citenamefont
  {Breckenridge}, \citenamefont {Gombos},\ and\ \citenamefont
  {Morgan}}]{Breckenridge1972}%
  \BibitemOpen
  \bibfield  {author} {\bibinfo {author} {\bibfnamefont {W.}~\bibnamefont
  {Breckenridge}}, \bibinfo {author} {\bibfnamefont {G.}~\bibnamefont
  {Gombos}},\ and\ \bibinfo {author} {\bibfnamefont {I.}~\bibnamefont
  {Morgan}},\ }\bibfield  {title} {\bibinfo {title} {The lipid composition of
  adult rat brain synaptosomal plasma membranes},\ }\href
  {https://doi.org/10.1016/0005-2736(72)90365-3} {\bibfield  {journal}
  {\bibinfo  {journal} {Biochim. Biophys. Acta, Biomembr.}\ }\textbf {\bibinfo
  {volume} {266}},\ \bibinfo {pages} {695 } (\bibinfo {year} {1972})},\ \Eprint
  {https://arxiv.org/abs/10.1016/0005-2736(72)90365-3}
  {10.1016/0005-2736(72)90365-3} \BibitemShut {NoStop}%
\bibitem [{\citenamefont {Drin}(2014)}]{Drin2014}%
  \BibitemOpen
  \bibfield  {author} {\bibinfo {author} {\bibfnamefont {G.}~\bibnamefont
  {Drin}},\ }\bibfield  {title} {\bibinfo {title} {Topological regulation of
  lipid balance in cells},\ }\href
  {https://doi.org/10.1146/annurev-biochem-060713-035307} {\bibfield  {journal}
  {\bibinfo  {journal} {Annu. Rev. Biochem.}\ }\textbf {\bibinfo {volume}
  {83}},\ \bibinfo {pages} {51} (\bibinfo {year} {2014})},\ \bibinfo {note}
  {pMID: 24606148},\ \Eprint
  {https://arxiv.org/abs/10.1146/annurev-biochem-060713-035307}
  {10.1146/annurev-biochem-060713-035307} \BibitemShut {NoStop}%
\bibitem [{\citenamefont {Israelachvili}(2011)}]{Israelachvili2011}%
  \BibitemOpen
  \bibfield  {author} {\bibinfo {author} {\bibfnamefont {J.~N.}\ \bibnamefont
  {Israelachvili}},\ }\href@noop {} {\emph {\bibinfo {title} {Intermolecular
  and Surface Forces}}},\ \bibinfo {edition} {3rd}\ ed.\ (\bibinfo  {publisher}
  {Academic Press},\ \bibinfo {year} {2011})\BibitemShut {NoStop}%
\bibitem [{\citenamefont {Li}\ \emph {et~al.}(2016)\citenamefont {Li},
  \citenamefont {Sun}, \citenamefont {Yap}, \citenamefont {Chen},\ and\
  \citenamefont {Qian}}]{Li2016}%
  \BibitemOpen
  \bibfield  {author} {\bibinfo {author} {\bibfnamefont {H.}~\bibnamefont
  {Li}}, \bibinfo {author} {\bibfnamefont {S.-r.}\ \bibnamefont {Sun}},
  \bibinfo {author} {\bibfnamefont {J.~Q.}\ \bibnamefont {Yap}}, \bibinfo
  {author} {\bibfnamefont {J.-h.}\ \bibnamefont {Chen}},\ and\ \bibinfo
  {author} {\bibfnamefont {Q.}~\bibnamefont {Qian}},\ }\bibfield  {title}
  {\bibinfo {title} {0.9{\%} saline is neither normal nor physiological},\
  }\href {https://doi.org/10.1631/jzus.B1500201} {\bibfield  {journal}
  {\bibinfo  {journal} {J. Zhejiang Univ., Sci., B}\ }\textbf {\bibinfo
  {volume} {17}},\ \bibinfo {pages} {181} (\bibinfo {year} {2016})},\ \Eprint
  {https://arxiv.org/abs/10.1631/jzus.B1500201} {10.1631/jzus.B1500201}
  \BibitemShut {NoStop}%
\bibitem [{\citenamefont {Tebaykin}\ \emph {et~al.}(2018)\citenamefont
  {Tebaykin}, \citenamefont {Tripathy}, \citenamefont {Binnion}, \citenamefont
  {Li}, \citenamefont {Gerkin},\ and\ \citenamefont {Pavlidis}}]{Tebaykin2018}%
  \BibitemOpen
  \bibfield  {author} {\bibinfo {author} {\bibfnamefont {D.}~\bibnamefont
  {Tebaykin}}, \bibinfo {author} {\bibfnamefont {S.~J.}\ \bibnamefont
  {Tripathy}}, \bibinfo {author} {\bibfnamefont {N.}~\bibnamefont {Binnion}},
  \bibinfo {author} {\bibfnamefont {B.}~\bibnamefont {Li}}, \bibinfo {author}
  {\bibfnamefont {R.~C.}\ \bibnamefont {Gerkin}},\ and\ \bibinfo {author}
  {\bibfnamefont {P.}~\bibnamefont {Pavlidis}},\ }\bibfield  {title} {\bibinfo
  {title} {Modeling sources of interlaboratory variability in
  electrophysiological properties of mammalian neurons},\ }\href
  {https://doi.org/10.1152/jn.00604.2017} {\bibfield  {journal} {\bibinfo
  {journal} {J. Neurophysiol.}\ }\textbf {\bibinfo {volume} {119}},\ \bibinfo
  {pages} {1329} (\bibinfo {year} {2018})},\ \bibinfo {note} {pMID: 29357465},\
  \Eprint {https://arxiv.org/abs/10.1152/jn.00604.2017} {10.1152/jn.00604.2017}
  \BibitemShut {NoStop}%
\bibitem [{\citenamefont {Lipowsky}(1995)}]{Lipowsky1995}%
  \BibitemOpen
  \bibfield  {author} {\bibinfo {author} {\bibfnamefont {R.}~\bibnamefont
  {Lipowsky}},\ }\bibinfo {title} {Generic interactions of flexible
  membranes},\ in\ \href@noop {} {\emph {\bibinfo {booktitle} {Handbook of
  Biological Physics}}},\ Vol.~\bibinfo {volume} {1b},\ \bibinfo {editor}
  {edited by\ \bibinfo {editor} {\bibfnamefont {R.}~\bibnamefont {Lipowsky}}\
  and\ \bibinfo {editor} {\bibfnamefont {E.}~\bibnamefont {Sackmann}}}\
  (\bibinfo  {publisher} {Elsevier},\ \bibinfo {address} {Amsterdam},\ \bibinfo
  {year} {1995})\ Chap.~\bibinfo {chapter} {11}, pp.\ \bibinfo {pages}
  {521--602}\BibitemShut {NoStop}%
\bibitem [{\citenamefont {Oosawa}(1968)}]{Oosawa1968}%
  \BibitemOpen
  \bibfield  {author} {\bibinfo {author} {\bibfnamefont {F.}~\bibnamefont
  {Oosawa}},\ }\bibfield  {title} {\bibinfo {title} {Interaction between
  parallel rodlike macroions},\ }\href
  {https://doi.org/10.1002/bip.1968.360061108} {\bibfield  {journal} {\bibinfo
  {journal} {Biopolymers}\ }\textbf {\bibinfo {volume} {6}},\ \bibinfo {pages}
  {1633} (\bibinfo {year} {1968})},\ \Eprint
  {https://arxiv.org/abs/10.1002/bip.1968.360061108}
  {10.1002/bip.1968.360061108} \BibitemShut {NoStop}%
\bibitem [{\citenamefont {Kekicheff}\ \emph {et~al.}(1993)\citenamefont
  {Kekicheff}, \citenamefont {Marcelja}, \citenamefont {Senden},\ and\
  \citenamefont {Shubin}}]{Kekicheff1993}%
  \BibitemOpen
  \bibfield  {author} {\bibinfo {author} {\bibfnamefont {P.}~\bibnamefont
  {Kekicheff}}, \bibinfo {author} {\bibfnamefont {S.}~\bibnamefont {Marcelja}},
  \bibinfo {author} {\bibfnamefont {T.~J.}\ \bibnamefont {Senden}},\ and\
  \bibinfo {author} {\bibfnamefont {V.~E.}\ \bibnamefont {Shubin}},\ }\bibfield
   {title} {\bibinfo {title} {Charge reversal seen in electrical double layer
  interaction of surfaces immersed in 2:1 calcium electrolyte},\ }\href
  {https://doi.org/10.1063/1.465906} {\bibfield  {journal} {\bibinfo  {journal}
  {J. Chem. Phys.}\ }\textbf {\bibinfo {volume} {99}},\ \bibinfo {pages} {6098}
  (\bibinfo {year} {1993})},\ \Eprint {https://arxiv.org/abs/10.1063/1.465906}
  {10.1063/1.465906} \BibitemShut {NoStop}%
\bibitem [{\citenamefont {Priel}\ \emph {et~al.}(2007)\citenamefont {Priel},
  \citenamefont {Gil}, \citenamefont {Moy}, \citenamefont {Magleby},\ and\
  \citenamefont {Silberberg}}]{Priel2007}%
  \BibitemOpen
  \bibfield  {author} {\bibinfo {author} {\bibfnamefont {A.}~\bibnamefont
  {Priel}}, \bibinfo {author} {\bibfnamefont {Z.}~\bibnamefont {Gil}}, \bibinfo
  {author} {\bibfnamefont {V.~T.}\ \bibnamefont {Moy}}, \bibinfo {author}
  {\bibfnamefont {K.~L.}\ \bibnamefont {Magleby}},\ and\ \bibinfo {author}
  {\bibfnamefont {S.~D.}\ \bibnamefont {Silberberg}},\ }\bibfield  {title}
  {\bibinfo {title} {Ionic requirements for membrane-glass adhesion and giga
  seal formation in patch-clamp recording},\ }\href
  {https://doi.org/10.1529/biophysj.106.099119} {\bibfield  {journal} {\bibinfo
   {journal} {Biophys. J.}\ }\textbf {\bibinfo {volume} {92}},\ \bibinfo
  {pages} {3893 } (\bibinfo {year} {2007})},\ \Eprint
  {https://arxiv.org/abs/10.1529/biophysj.106.099119}
  {10.1529/biophysj.106.099119} \BibitemShut {NoStop}%
\bibitem [{\citenamefont {France-Lanord}\ and\ \citenamefont
  {Grossman}(2019)}]{FranceLanord2019}%
  \BibitemOpen
  \bibfield  {author} {\bibinfo {author} {\bibfnamefont {A.}~\bibnamefont
  {France-Lanord}}\ and\ \bibinfo {author} {\bibfnamefont {J.~C.}\ \bibnamefont
  {Grossman}},\ }\bibfield  {title} {\bibinfo {title} {Correlations from ion
  pairing and the {Nernst}-{Einstein} equation},\ }\href
  {https://doi.org/10.1103/PhysRevLett.122.136001} {\bibfield  {journal}
  {\bibinfo  {journal} {Phys. Rev. Lett.}\ }\textbf {\bibinfo {volume} {122}},\
  \bibinfo {pages} {136001} (\bibinfo {year} {2019})},\ \Eprint
  {https://arxiv.org/abs/10.1103/PhysRevLett.122.136001}
  {10.1103/PhysRevLett.122.136001} \BibitemShut {NoStop}%
\bibitem [{\citenamefont {Leckband}\ and\ \citenamefont
  {Israelachvili}(2001)}]{Leckband2001}%
  \BibitemOpen
  \bibfield  {author} {\bibinfo {author} {\bibfnamefont {D.}~\bibnamefont
  {Leckband}}\ and\ \bibinfo {author} {\bibfnamefont {J.}~\bibnamefont
  {Israelachvili}},\ }\bibfield  {title} {\bibinfo {title} {Intermolecular
  forces in biology},\ }\href {https://doi.org/10.1017/S0033583501003687}
  {\bibfield  {journal} {\bibinfo  {journal} {Q. Rev. Biophys.}\ }\textbf
  {\bibinfo {volume} {34}},\ \bibinfo {pages} {105–267} (\bibinfo {year}
  {2001})},\ \Eprint {https://arxiv.org/abs/10.1017/S0033583501003687}
  {10.1017/S0033583501003687} \BibitemShut {NoStop}%
\bibitem [{\citenamefont {Casimir}\ and\ \citenamefont
  {Polder}(1948)}]{Casimir1948}%
  \BibitemOpen
  \bibfield  {author} {\bibinfo {author} {\bibfnamefont {H.~B.~G.}\
  \bibnamefont {Casimir}}\ and\ \bibinfo {author} {\bibfnamefont
  {D.}~\bibnamefont {Polder}},\ }\bibfield  {title} {\bibinfo {title} {The
  influence of retardation on the {L}ondon-van der {W}aals forces},\ }\href
  {https://doi.org/10.1103/PhysRev.73.360} {\bibfield  {journal} {\bibinfo
  {journal} {Phys. Rev.}\ }\textbf {\bibinfo {volume} {73}},\ \bibinfo {pages}
  {360} (\bibinfo {year} {1948})},\ \Eprint
  {https://arxiv.org/abs/10.1103/PhysRev.73.360} {10.1103/PhysRev.73.360}
  \BibitemShut {NoStop}%
\bibitem [{\citenamefont {Matsubara}(1955)}]{Matsubara1955}%
  \BibitemOpen
  \bibfield  {author} {\bibinfo {author} {\bibfnamefont {T.}~\bibnamefont
  {Matsubara}},\ }\bibfield  {title} {\bibinfo {title} {A new approach to
  quantum-statistical mechanics},\ }\href {https://doi.org/10.1143/PTP.14.351}
  {\bibfield  {journal} {\bibinfo  {journal} {Prog. Theor. Phys.}\ }\textbf
  {\bibinfo {volume} {14}},\ \bibinfo {pages} {351} (\bibinfo {year} {1955})},\
  \Eprint
  {https://arxiv.org/abs/https://academic.oup.com/ptp/article-pdf/14/4/351/5286981/14-4-351.pdf}
  {https://academic.oup.com/ptp/article-pdf/14/4/351/5286981/14-4-351.pdf}
  \BibitemShut {NoStop}%
\bibitem [{\citenamefont {Dzyaloshinskii}\ \emph {et~al.}(1961)\citenamefont
  {Dzyaloshinskii}, \citenamefont {Lifshitz},\ and\ \citenamefont
  {Pitaevskii}}]{Dzyaloshinskii1961}%
  \BibitemOpen
  \bibfield  {author} {\bibinfo {author} {\bibfnamefont {I.~E.}\ \bibnamefont
  {Dzyaloshinskii}}, \bibinfo {author} {\bibfnamefont {E.~M.}\ \bibnamefont
  {Lifshitz}},\ and\ \bibinfo {author} {\bibfnamefont {L.~P.}\ \bibnamefont
  {Pitaevskii}},\ }\bibfield  {title} {\bibinfo {title} {General theory of van
  der {Waals} forces},\ }\href
  {https://doi.org/10.1070/pu1961v004n02abeh003330} {\bibfield  {journal}
  {\bibinfo  {journal} {Phys.-Usp.}\ }\textbf {\bibinfo {volume} {4}},\
  \bibinfo {pages} {153} (\bibinfo {year} {1961})},\ \Eprint
  {https://arxiv.org/abs/10.1070/pu1961v004n02abeh003330}
  {10.1070/pu1961v004n02abeh003330} \BibitemShut {NoStop}%
\bibitem [{\citenamefont {Borosilicates}(2019)}]{GlassIndustryData}%
  \BibitemOpen
  \bibfield  {author} {\bibinfo {author} {\bibnamefont {Borosilicates}},\
  }\href@noop {} {} (\bibinfo {year} {2019}),\ \bibinfo {note} {{C}orning 7052:
  $\varepsilon_{r}=5.1$, $n=1.484$ (Corning), 7740: $\varepsilon_{r}=5.0$,
  $n=1.473$ (GlassFab); $\varepsilon_{r}=4.6$, $n=1.474$ (Thermofisher). Schott
  8250: $\varepsilon_{r}=4.9$, $n=1.487$ (WPI)}\BibitemShut {NoStop}%
\bibitem [{\citenamefont {B{\"u}ldt}\ \emph {et~al.}(1978)\citenamefont
  {B{\"u}ldt}, \citenamefont {Gally}, \citenamefont {Seelig}, \citenamefont
  {Seelig},\ and\ \citenamefont {Zaccai}}]{1978Buldt}%
  \BibitemOpen
  \bibfield  {author} {\bibinfo {author} {\bibfnamefont {G.}~\bibnamefont
  {B{\"u}ldt}}, \bibinfo {author} {\bibfnamefont {H.}~\bibnamefont {Gally}},
  \bibinfo {author} {\bibfnamefont {A.}~\bibnamefont {Seelig}}, \bibinfo
  {author} {\bibfnamefont {J.}~\bibnamefont {Seelig}},\ and\ \bibinfo {author}
  {\bibfnamefont {G.}~\bibnamefont {Zaccai}},\ }\bibfield  {title} {\bibinfo
  {title} {Neutron diffraction studies on selectively deuterated phospholipid
  bilayers},\ }\href {https://doi.org/10.1038/271182a0} {\bibfield  {journal}
  {\bibinfo  {journal} {Nature}\ }\textbf {\bibinfo {volume} {271}},\ \bibinfo
  {pages} {182} (\bibinfo {year} {1978})}\BibitemShut {NoStop}%
\bibitem [{\citenamefont {Lucas}\ \emph {et~al.}(2012)\citenamefont {Lucas},
  \citenamefont {Bauer}, \citenamefont {Davis},\ and\ \citenamefont
  {Patel}}]{Lucas2012}%
  \BibitemOpen
  \bibfield  {author} {\bibinfo {author} {\bibfnamefont {T.~R.}\ \bibnamefont
  {Lucas}}, \bibinfo {author} {\bibfnamefont {B.~A.}\ \bibnamefont {Bauer}},
  \bibinfo {author} {\bibfnamefont {J.~E.}\ \bibnamefont {Davis}},\ and\
  \bibinfo {author} {\bibfnamefont {S.}~\bibnamefont {Patel}},\ }\bibfield
  {title} {\bibinfo {title} {Molecular dynamics simulation of hydrated dppc
  monolayers using charge equilibration force fields},\ }\href
  {https://doi.org/10.1002/jcc.21927} {\bibfield  {journal} {\bibinfo
  {journal} {J. Comput. Chem.}\ }\textbf {\bibinfo {volume} {33}},\ \bibinfo
  {pages} {141} (\bibinfo {year} {2012})},\ \Eprint
  {https://arxiv.org/abs/10.1002/jcc.21927} {10.1002/jcc.21927} \BibitemShut
  {NoStop}%
\bibitem [{\citenamefont {Stern}\ and\ \citenamefont
  {Feller}(2003)}]{Stern2003}%
  \BibitemOpen
  \bibfield  {author} {\bibinfo {author} {\bibfnamefont {H.~A.}\ \bibnamefont
  {Stern}}\ and\ \bibinfo {author} {\bibfnamefont {S.~E.}\ \bibnamefont
  {Feller}},\ }\bibfield  {title} {\bibinfo {title} {Calculation of the
  dielectric permittivity profile for a nonuniform system: Application to a
  lipid bilayer simulation},\ }\href {https://doi.org/10.1063/1.1537244}
  {\bibfield  {journal} {\bibinfo  {journal} {J. Chem. Phys.}\ }\textbf
  {\bibinfo {volume} {118}},\ \bibinfo {pages} {3401} (\bibinfo {year}
  {2003})},\ \Eprint {https://arxiv.org/abs/10.1063/1.1537244}
  {10.1063/1.1537244} \BibitemShut {NoStop}%
\bibitem [{\citenamefont {Nymeyer}\ and\ \citenamefont
  {Zhou}(2008)}]{Nymeyer2008}%
  \BibitemOpen
  \bibfield  {author} {\bibinfo {author} {\bibfnamefont {H.}~\bibnamefont
  {Nymeyer}}\ and\ \bibinfo {author} {\bibfnamefont {H.-X.}\ \bibnamefont
  {Zhou}},\ }\bibfield  {title} {\bibinfo {title} {A method to determine
  dielectric constants in nonhomogeneous systems: Application to biological
  membranes},\ }\href {https://doi.org/10.1529/biophysj.107.117770} {\bibfield
  {journal} {\bibinfo  {journal} {Biophys. J.}\ }\textbf {\bibinfo {volume}
  {94}},\ \bibinfo {pages} {1185 } (\bibinfo {year} {2008})},\ \Eprint
  {https://arxiv.org/abs/10.1529/biophysj.107.117770}
  {10.1529/biophysj.107.117770} \BibitemShut {NoStop}%
\bibitem [{\citenamefont {Spector}\ \emph {et~al.}(1996)\citenamefont
  {Spector}, \citenamefont {Easwaran}, \citenamefont {Jyothi}, \citenamefont
  {Selinger}, \citenamefont {Singh},\ and\ \citenamefont
  {Schnur}}]{Spector1996}%
  \BibitemOpen
  \bibfield  {author} {\bibinfo {author} {\bibfnamefont {M.~S.}\ \bibnamefont
  {Spector}}, \bibinfo {author} {\bibfnamefont {K.~R.~K.}\ \bibnamefont
  {Easwaran}}, \bibinfo {author} {\bibfnamefont {G.}~\bibnamefont {Jyothi}},
  \bibinfo {author} {\bibfnamefont {J.~V.}\ \bibnamefont {Selinger}}, \bibinfo
  {author} {\bibfnamefont {A.}~\bibnamefont {Singh}},\ and\ \bibinfo {author}
  {\bibfnamefont {J.~M.}\ \bibnamefont {Schnur}},\ }\bibfield  {title}
  {\bibinfo {title} {Chiral molecular self-assembly of phospholipid tubules: A
  circular dichroism study},\ }\href {https://doi.org/10.1073/pnas.93.23.12943}
  {\bibfield  {journal} {\bibinfo  {journal} {Proc. Nat. Acad. Sci. U.S.A.}\
  }\textbf {\bibinfo {volume} {93}},\ \bibinfo {pages} {12943} (\bibinfo {year}
  {1996})},\ \Eprint {https://arxiv.org/abs/10.1073/pnas.93.23.12943}
  {10.1073/pnas.93.23.12943} \BibitemShut {NoStop}%
\bibitem [{\citenamefont {Gramse}\ \emph {et~al.}(2013)\citenamefont {Gramse},
  \citenamefont {Dols-Perez}, \citenamefont {Edwards}, \citenamefont
  {Fumagalli},\ and\ \citenamefont {Gomila}}]{Gramse2013}%
  \BibitemOpen
  \bibfield  {author} {\bibinfo {author} {\bibfnamefont {G.}~\bibnamefont
  {Gramse}}, \bibinfo {author} {\bibfnamefont {A.}~\bibnamefont {Dols-Perez}},
  \bibinfo {author} {\bibfnamefont {M.}~\bibnamefont {Edwards}}, \bibinfo
  {author} {\bibfnamefont {L.}~\bibnamefont {Fumagalli}},\ and\ \bibinfo
  {author} {\bibfnamefont {G.}~\bibnamefont {Gomila}},\ }\bibfield  {title}
  {\bibinfo {title} {Nanoscale measurement of the dielectric constant of
  supported lipid bilayers in aqueous solutions with electrostatic force
  microscopy},\ }\href {https://doi.org/10.1016/j.bpj.2013.02.011} {\bibfield
  {journal} {\bibinfo  {journal} {Biophys. J.}\ }\textbf {\bibinfo {volume}
  {104}},\ \bibinfo {pages} {1257 } (\bibinfo {year} {2013})},\ \Eprint
  {https://arxiv.org/abs/10.1016/j.bpj.2013.02.011} {10.1016/j.bpj.2013.02.011}
  \BibitemShut {NoStop}%
\bibitem [{\citenamefont {Maxwell}(1891)}]{Maxwell1891}%
  \BibitemOpen
  \bibfield  {author} {\bibinfo {author} {\bibfnamefont {J.~C.}\ \bibnamefont
  {Maxwell}},\ }\href@noop {} {\emph {\bibinfo {title} {A Treatise on
  Electricity and Magnetism}}},\ \bibinfo {edition} {3rd}\ ed.\ (\bibinfo
  {publisher} {Clarendon Press},\ \bibinfo {address} {London},\ \bibinfo {year}
  {1891})\BibitemShut {NoStop}%
\bibitem [{\citenamefont {Gabriel}\ \emph {et~al.}(1996)\citenamefont
  {Gabriel}, \citenamefont {Lau},\ and\ \citenamefont {Gabriel}}]{Gabriel1996}%
  \BibitemOpen
  \bibfield  {author} {\bibinfo {author} {\bibfnamefont {S.}~\bibnamefont
  {Gabriel}}, \bibinfo {author} {\bibfnamefont {R.~W.}\ \bibnamefont {Lau}},\
  and\ \bibinfo {author} {\bibfnamefont {C.}~\bibnamefont {Gabriel}},\
  }\bibfield  {title} {\bibinfo {title} {The dielectric properties of
  biological tissues: {II}. measurements in the frequency range 10 {Hz} to 20
  {GHz}},\ }\href {https://doi.org/10.1088/0031-9155/41/11/002} {\bibfield
  {journal} {\bibinfo  {journal} {Phys. Med. Biol.}\ }\textbf {\bibinfo
  {volume} {41}},\ \bibinfo {pages} {2251} (\bibinfo {year} {1996})},\ \Eprint
  {https://arxiv.org/abs/10.1088/0031-9155/41/11/002}
  {10.1088/0031-9155/41/11/002} \BibitemShut {NoStop}%
\bibitem [{\citenamefont {Koenig}\ \emph {et~al.}(1996)\citenamefont {Koenig},
  \citenamefont {Krueger}, \citenamefont {Orts}, \citenamefont {Majkrzak},
  \citenamefont {Berk}, \citenamefont {Silverton},\ and\ \citenamefont
  {Gawrisch}}]{1996Koenig}%
  \BibitemOpen
  \bibfield  {author} {\bibinfo {author} {\bibfnamefont {B.~W.}\ \bibnamefont
  {Koenig}}, \bibinfo {author} {\bibfnamefont {S.}~\bibnamefont {Krueger}},
  \bibinfo {author} {\bibfnamefont {W.}~\bibnamefont {Orts}}, \bibinfo {author}
  {\bibfnamefont {C.~F.}\ \bibnamefont {Majkrzak}}, \bibinfo {author}
  {\bibfnamefont {N.~F.}\ \bibnamefont {Berk}}, \bibinfo {author}
  {\bibfnamefont {J.}~\bibnamefont {Silverton}},\ and\ \bibinfo {author}
  {\bibfnamefont {K.}~\bibnamefont {Gawrisch}},\ }\bibfield  {title} {\bibinfo
  {title} {Neutron reflectivity and atomic force microscopy studies of a lipid
  bilayer in water adsorbed to the surface of a silicon single crystal},\
  }\href {https://doi.org/10.1021/la950580r} {\bibfield  {journal} {\bibinfo
  {journal} {Langmuir}\ }\textbf {\bibinfo {volume} {12}},\ \bibinfo {pages}
  {1343} (\bibinfo {year} {1996})}\BibitemShut {NoStop}%
\bibitem [{\citenamefont {Ardhammar}\ \emph {et~al.}(2002)\citenamefont
  {Ardhammar}, \citenamefont {Lincoln},\ and\ \citenamefont
  {Nord{\'e}n}}]{Ardhammar2002}%
  \BibitemOpen
  \bibfield  {author} {\bibinfo {author} {\bibfnamefont {M.}~\bibnamefont
  {Ardhammar}}, \bibinfo {author} {\bibfnamefont {P.}~\bibnamefont {Lincoln}},\
  and\ \bibinfo {author} {\bibfnamefont {B.}~\bibnamefont {Nord{\'e}n}},\
  }\bibfield  {title} {\bibinfo {title} {Invisible liposomes: Refractive index
  matching with sucrose enables flow dichroism assessment of peptide
  orientation in lipid vesicle membrane},\ }\href
  {https://doi.org/10.1073/pnas.192583499} {\bibfield  {journal} {\bibinfo
  {journal} {Proc. Nat. Acad. Sci. U.S.A.}\ }\textbf {\bibinfo {volume} {99}},\
  \bibinfo {pages} {15313} (\bibinfo {year} {2002})},\ \Eprint
  {https://arxiv.org/abs/10.1073/pnas.192583499} {10.1073/pnas.192583499}
  \BibitemShut {NoStop}%
\bibitem [{\citenamefont {Granqvist}\ \emph {et~al.}(2014)\citenamefont
  {Granqvist}, \citenamefont {Yliperttula}, \citenamefont {V{\"a}lim{\"a}ki},
  \citenamefont {Pulkkinen}, \citenamefont {Tenhu},\ and\ \citenamefont
  {Viitala}}]{2014Granqvist}%
  \BibitemOpen
  \bibfield  {author} {\bibinfo {author} {\bibfnamefont {N.}~\bibnamefont
  {Granqvist}}, \bibinfo {author} {\bibfnamefont {M.}~\bibnamefont
  {Yliperttula}}, \bibinfo {author} {\bibfnamefont {S.}~\bibnamefont
  {V{\"a}lim{\"a}ki}}, \bibinfo {author} {\bibfnamefont {P.}~\bibnamefont
  {Pulkkinen}}, \bibinfo {author} {\bibfnamefont {H.}~\bibnamefont {Tenhu}},\
  and\ \bibinfo {author} {\bibfnamefont {T.}~\bibnamefont {Viitala}},\
  }\bibfield  {title} {\bibinfo {title} {Control of the morphology of lipid
  layers by substrate surface chemistry},\ }\href
  {https://doi.org/10.1021/la4046622} {\bibfield  {journal} {\bibinfo
  {journal} {Langmuir}\ }\textbf {\bibinfo {volume} {30}},\ \bibinfo {pages}
  {2799} (\bibinfo {year} {2014})}\BibitemShut {NoStop}%
\bibitem [{\citenamefont {Parkkila}\ \emph {et~al.}(2018)\citenamefont
  {Parkkila}, \citenamefont {Elderdfi}, \citenamefont {Bunker},\ and\
  \citenamefont {Viitala}}]{2018Parkkila}%
  \BibitemOpen
  \bibfield  {author} {\bibinfo {author} {\bibfnamefont {P.}~\bibnamefont
  {Parkkila}}, \bibinfo {author} {\bibfnamefont {M.}~\bibnamefont {Elderdfi}},
  \bibinfo {author} {\bibfnamefont {A.}~\bibnamefont {Bunker}},\ and\ \bibinfo
  {author} {\bibfnamefont {T.}~\bibnamefont {Viitala}},\ }\bibfield  {title}
  {\bibinfo {title} {Biophysical characterization of supported lipid bilayers
  using parallel dual-wavelength surface plasmon resonance and quartz crystal
  microbalance measurements},\ }\href
  {https://doi.org/10.1021/acs.langmuir.8b01259} {\bibfield  {journal}
  {\bibinfo  {journal} {Langmuir}\ }\textbf {\bibinfo {volume} {34}},\ \bibinfo
  {pages} {8081} (\bibinfo {year} {2018})}\BibitemShut {NoStop}%
\bibitem [{\citenamefont {Sch{\"o}nherr}\ \emph {et~al.}(2004)\citenamefont
  {Sch{\"o}nherr}, \citenamefont {Johnson}, \citenamefont {Lenz}, \citenamefont
  {Frank},\ and\ \citenamefont {Boxer}}]{2004Schonherr}%
  \BibitemOpen
  \bibfield  {author} {\bibinfo {author} {\bibfnamefont {H.}~\bibnamefont
  {Sch{\"o}nherr}}, \bibinfo {author} {\bibfnamefont {J.~M.}\ \bibnamefont
  {Johnson}}, \bibinfo {author} {\bibfnamefont {P.}~\bibnamefont {Lenz}},
  \bibinfo {author} {\bibfnamefont {C.~W.}\ \bibnamefont {Frank}},\ and\
  \bibinfo {author} {\bibfnamefont {S.~G.}\ \bibnamefont {Boxer}},\ }\bibfield
  {title} {\bibinfo {title} {Vesicle adsorption and lipid bilayer formation on
  glass studied by atomic force microscopy},\ }\href
  {https://doi.org/10.1021/la049302v} {\bibfield  {journal} {\bibinfo
  {journal} {Langmuir}\ }\textbf {\bibinfo {volume} {20}},\ \bibinfo {pages}
  {11600} (\bibinfo {year} {2004})}\BibitemShut {NoStop}%
\bibitem [{\citenamefont {Kurniawan}\ \emph {et~al.}(2018)\citenamefont
  {Kurniawan}, \citenamefont {Ventrici~de Souza}, \citenamefont {Dang},
  \citenamefont {Liu},\ and\ \citenamefont {Kuhl}}]{2018Kurniawan}%
  \BibitemOpen
  \bibfield  {author} {\bibinfo {author} {\bibfnamefont {J.}~\bibnamefont
  {Kurniawan}}, \bibinfo {author} {\bibfnamefont {J.~F.}\ \bibnamefont
  {Ventrici~de Souza}}, \bibinfo {author} {\bibfnamefont {A.~T.}\ \bibnamefont
  {Dang}}, \bibinfo {author} {\bibfnamefont {G.-y.}\ \bibnamefont {Liu}},\ and\
  \bibinfo {author} {\bibfnamefont {T.~L.}\ \bibnamefont {Kuhl}},\ }\bibfield
  {title} {\bibinfo {title} {Preparation and characterization of
  solid-supported lipid bilayers formed by langmuir--blodgett deposition: A
  tutorial},\ }\href {https://doi.org/10.1021/acs.langmuir.8b03504} {\bibfield
  {journal} {\bibinfo  {journal} {Langmuir}\ }\textbf {\bibinfo {volume}
  {34}},\ \bibinfo {pages} {15622} (\bibinfo {year} {2018})}\BibitemShut
  {NoStop}%
\bibitem [{\citenamefont {Biswas}\ \emph {et~al.}(2018)\citenamefont {Biswas},
  \citenamefont {Jackman}, \citenamefont {Park}, \citenamefont {Groves},\ and\
  \citenamefont {Cho}}]{2018Biswas}%
  \BibitemOpen
  \bibfield  {author} {\bibinfo {author} {\bibfnamefont {K.~H.}\ \bibnamefont
  {Biswas}}, \bibinfo {author} {\bibfnamefont {J.~A.}\ \bibnamefont {Jackman}},
  \bibinfo {author} {\bibfnamefont {J.~H.}\ \bibnamefont {Park}}, \bibinfo
  {author} {\bibfnamefont {J.~T.}\ \bibnamefont {Groves}},\ and\ \bibinfo
  {author} {\bibfnamefont {N.-J.}\ \bibnamefont {Cho}},\ }\bibfield  {title}
  {\bibinfo {title} {Interfacial forces dictate the pathway of phospholipid
  vesicle adsorption onto silicon dioxide surfaces},\ }\href
  {https://doi.org/10.1021/acs.langmuir.7b03799} {\bibfield  {journal}
  {\bibinfo  {journal} {Langmuir}\ }\textbf {\bibinfo {volume} {34}},\ \bibinfo
  {pages} {1775} (\bibinfo {year} {2018})}\BibitemShut {NoStop}%
\bibitem [{\citenamefont {Richter}\ \emph {et~al.}(2006)\citenamefont
  {Richter}, \citenamefont {B{\'e}rat},\ and\ \citenamefont
  {Brisson}}]{2006Richter}%
  \BibitemOpen
  \bibfield  {author} {\bibinfo {author} {\bibfnamefont {R.~P.}\ \bibnamefont
  {Richter}}, \bibinfo {author} {\bibfnamefont {R.}~\bibnamefont {B{\'e}rat}},\
  and\ \bibinfo {author} {\bibfnamefont {A.~R.}\ \bibnamefont {Brisson}},\
  }\bibfield  {title} {\bibinfo {title} {Formation of solid-supported lipid
  bilayers: an integrated view},\ }\href {https://doi.org/10.1021/la052687c}
  {\bibfield  {journal} {\bibinfo  {journal} {Langmuir}\ }\textbf {\bibinfo
  {volume} {22}},\ \bibinfo {pages} {3497} (\bibinfo {year}
  {2006})}\BibitemShut {NoStop}%
\bibitem [{\citenamefont {Richter}\ \emph {et~al.}(2003)\citenamefont
  {Richter}, \citenamefont {Mukhopadhyay},\ and\ \citenamefont
  {Brisson}}]{2003Richter}%
  \BibitemOpen
  \bibfield  {author} {\bibinfo {author} {\bibfnamefont {R.}~\bibnamefont
  {Richter}}, \bibinfo {author} {\bibfnamefont {A.}~\bibnamefont
  {Mukhopadhyay}},\ and\ \bibinfo {author} {\bibfnamefont {A.}~\bibnamefont
  {Brisson}},\ }\bibfield  {title} {\bibinfo {title} {Pathways of lipid vesicle
  deposition on solid surfaces: a combined {QCM-D} and {AFM} study},\ }\href
  {https://doi.org/10.1016/S0006-3495(03)74722-5} {\bibfield  {journal}
  {\bibinfo  {journal} {Biophys. J.}\ }\textbf {\bibinfo {volume} {85}},\
  \bibinfo {pages} {3035} (\bibinfo {year} {2003})}\BibitemShut {NoStop}%
\bibitem [{\citenamefont {Brian}\ and\ \citenamefont
  {McConnell}(1984)}]{1984Brian}%
  \BibitemOpen
  \bibfield  {author} {\bibinfo {author} {\bibfnamefont {A.~A.}\ \bibnamefont
  {Brian}}\ and\ \bibinfo {author} {\bibfnamefont {H.~M.}\ \bibnamefont
  {McConnell}},\ }\bibfield  {title} {\bibinfo {title} {Allogeneic stimulation
  of cytotoxic t cells by supported planar membranes},\ }\href@noop {}
  {\bibfield  {journal} {\bibinfo  {journal} {Proc. Nat. Acad. Sci. U.S.A.}\
  }\textbf {\bibinfo {volume} {81}},\ \bibinfo {pages} {6159} (\bibinfo {year}
  {1984})}\BibitemShut {NoStop}%
\bibitem [{\citenamefont {McConnell}\ \emph {et~al.}(1986)\citenamefont
  {McConnell}, \citenamefont {Watts}, \citenamefont {Weis},\ and\ \citenamefont
  {Brian}}]{1986McConnell}%
  \BibitemOpen
  \bibfield  {author} {\bibinfo {author} {\bibfnamefont {H.}~\bibnamefont
  {McConnell}}, \bibinfo {author} {\bibfnamefont {T.}~\bibnamefont {Watts}},
  \bibinfo {author} {\bibfnamefont {R.}~\bibnamefont {Weis}},\ and\ \bibinfo
  {author} {\bibfnamefont {A.}~\bibnamefont {Brian}},\ }\bibfield  {title}
  {\bibinfo {title} {Supported planar membranes in studies of cell-cell
  recognition in the immune system},\ }\href
  {https://doi.org/https://doi.org/10.1016/0304-4157(86)90016-X} {\bibfield
  {journal} {\bibinfo  {journal} {Biochim. Biophys. Acta}\ }\textbf {\bibinfo
  {volume} {864}},\ \bibinfo {pages} {95 } (\bibinfo {year}
  {1986})}\BibitemShut {NoStop}%
\bibitem [{\citenamefont {Slavchov}\ \emph {et~al.}(2014)\citenamefont
  {Slavchov}, \citenamefont {Nomura}, \citenamefont {Martinac}, \citenamefont
  {Sokabe},\ and\ \citenamefont {Sachs}}]{Slavchov2014}%
  \BibitemOpen
  \bibfield  {author} {\bibinfo {author} {\bibfnamefont {R.~I.}\ \bibnamefont
  {Slavchov}}, \bibinfo {author} {\bibfnamefont {T.}~\bibnamefont {Nomura}},
  \bibinfo {author} {\bibfnamefont {B.}~\bibnamefont {Martinac}}, \bibinfo
  {author} {\bibfnamefont {M.}~\bibnamefont {Sokabe}},\ and\ \bibinfo {author}
  {\bibfnamefont {F.}~\bibnamefont {Sachs}},\ }\bibfield  {title} {\bibinfo
  {title} {Gigaseal mechanics: Creep of the gigaseal under the action of
  pressure, adhesion, and voltage},\ }\href {https://doi.org/10.1021/jp506965v}
  {\bibfield  {journal} {\bibinfo  {journal} {J. Phys. Chem. B}\ }\textbf
  {\bibinfo {volume} {118}},\ \bibinfo {pages} {12660} (\bibinfo {year}
  {2014})},\ \bibinfo {note} {pMID: 25295693},\ \Eprint
  {https://arxiv.org/abs/10.1021/jp506965v} {10.1021/jp506965v} \BibitemShut
  {NoStop}%
\bibitem [{\citenamefont {Kolb}\ \emph {et~al.}(2016)\citenamefont {Kolb},
  \citenamefont {Stoy}, \citenamefont {Rousseau}, \citenamefont {Moody},
  \citenamefont {Jenkins},\ and\ \citenamefont {Forest}}]{Kolb2016}%
  \BibitemOpen
  \bibfield  {author} {\bibinfo {author} {\bibfnamefont {I.}~\bibnamefont
  {Kolb}}, \bibinfo {author} {\bibfnamefont {W.}~\bibnamefont {Stoy}}, \bibinfo
  {author} {\bibfnamefont {E.}~\bibnamefont {Rousseau}}, \bibinfo {author}
  {\bibfnamefont {O.}~\bibnamefont {Moody}}, \bibinfo {author} {\bibfnamefont
  {A.}~\bibnamefont {Jenkins}},\ and\ \bibinfo {author} {\bibfnamefont
  {C.}~\bibnamefont {Forest}},\ }\bibfield  {title} {\bibinfo {title} {Cleaning
  patch-clamp pipettes for immediate reuse},\ }\href
  {https://doi.org/10.1038/srep35001} {\bibfield  {journal} {\bibinfo
  {journal} {Sci. Rep.}\ }\textbf {\bibinfo {volume} {6}},\ \bibinfo {pages}
  {35001} (\bibinfo {year} {2016})},\ \Eprint
  {https://arxiv.org/abs/10.1038/srep35001} {10.1038/srep35001} \BibitemShut
  {NoStop}%
\bibitem [{\citenamefont {Needham}\ and\ \citenamefont
  {Nunn}(1990)}]{Needham1990}%
  \BibitemOpen
  \bibfield  {author} {\bibinfo {author} {\bibfnamefont {D.}~\bibnamefont
  {Needham}}\ and\ \bibinfo {author} {\bibfnamefont {R.}~\bibnamefont {Nunn}},\
  }\bibfield  {title} {\bibinfo {title} {Elastic deformation and failure of
  lipid bilayer membranes containing cholesterol},\ }\href
  {https://doi.org/10.1016/S0006-3495(90)82444-9} {\bibfield  {journal}
  {\bibinfo  {journal} {Biophys. J.}\ }\textbf {\bibinfo {volume} {58}},\
  \bibinfo {pages} {997 } (\bibinfo {year} {1990})},\ \Eprint
  {https://arxiv.org/abs/10.1016/S0006-3495(90)82444-9}
  {10.1016/S0006-3495(90)82444-9} \BibitemShut {NoStop}%
\bibitem [{\citenamefont {Dimova}(2014)}]{Dimova2014}%
  \BibitemOpen
  \bibfield  {author} {\bibinfo {author} {\bibfnamefont {R.}~\bibnamefont
  {Dimova}},\ }\bibfield  {title} {\bibinfo {title} {Recent developments in the
  field of bending rigidity measurements on membranes},\ }\href
  {https://doi.org/10.1016/j.cis.2014.03.003} {\bibfield  {journal} {\bibinfo
  {journal} {J. Colloid Interface Sci.}\ }\textbf {\bibinfo {volume} {208}},\
  \bibinfo {pages} {225 } (\bibinfo {year} {2014})},\ \Eprint
  {https://arxiv.org/abs/10.1016/j.cis.2014.03.003} {10.1016/j.cis.2014.03.003}
  \BibitemShut {NoStop}%
\bibitem [{\citenamefont {Newcomb}\ \emph {et~al.}(2017)\citenamefont
  {Newcomb}, \citenamefont {Fontana}, \citenamefont {Winkler}, \citenamefont
  {Cheng}, \citenamefont {Heymann},\ and\ \citenamefont
  {Steven}}]{2017Newcomb}%
  \BibitemOpen
  \bibfield  {author} {\bibinfo {author} {\bibfnamefont {W.~W.}\ \bibnamefont
  {Newcomb}}, \bibinfo {author} {\bibfnamefont {J.}~\bibnamefont {Fontana}},
  \bibinfo {author} {\bibfnamefont {D.~C.}\ \bibnamefont {Winkler}}, \bibinfo
  {author} {\bibfnamefont {N.}~\bibnamefont {Cheng}}, \bibinfo {author}
  {\bibfnamefont {J.~B.}\ \bibnamefont {Heymann}},\ and\ \bibinfo {author}
  {\bibfnamefont {A.~C.}\ \bibnamefont {Steven}},\ }\bibfield  {title}
  {\bibinfo {title} {The primary enveloped virion of herpes simplex virus 1:
  its role in nuclear egress},\ }\bibfield  {journal} {\bibinfo  {journal}
  {MBio}\ }\textbf {\bibinfo {volume} {8}},\ \href
  {https://doi.org/10.1128/mBio.00825-17} {10.1128/mBio.00825-17} (\bibinfo
  {year} {2017})\BibitemShut {NoStop}%
\bibitem [{\citenamefont {White}\ \emph {et~al.}(2016)\citenamefont {White},
  \citenamefont {Estrada}, \citenamefont {Flood}, \citenamefont {Mahmood},
  \citenamefont {Dhere},\ and\ \citenamefont {Chen}}]{2016White}%
  \BibitemOpen
  \bibfield  {author} {\bibinfo {author} {\bibfnamefont {J.~A.}\ \bibnamefont
  {White}}, \bibinfo {author} {\bibfnamefont {M.}~\bibnamefont {Estrada}},
  \bibinfo {author} {\bibfnamefont {E.~A.}\ \bibnamefont {Flood}}, \bibinfo
  {author} {\bibfnamefont {K.}~\bibnamefont {Mahmood}}, \bibinfo {author}
  {\bibfnamefont {R.}~\bibnamefont {Dhere}},\ and\ \bibinfo {author}
  {\bibfnamefont {D.}~\bibnamefont {Chen}},\ }\bibfield  {title} {\bibinfo
  {title} {Development of a stable liquid formulation of live attenuated
  influenza vaccine},\ }\href {https://doi.org/10.1016/j.vaccine.2016.04.074}
  {\bibfield  {journal} {\bibinfo  {journal} {Vaccine}\ }\textbf {\bibinfo
  {volume} {34}},\ \bibinfo {pages} {3676} (\bibinfo {year}
  {2016})}\BibitemShut {NoStop}%
\bibitem [{\citenamefont {Betz}\ \emph {et~al.}(2011)\citenamefont {Betz},
  \citenamefont {Koch}, \citenamefont {Lu}, \citenamefont {Franze},\ and\
  \citenamefont {K{\"a}s}}]{Betz2011}%
  \BibitemOpen
  \bibfield  {author} {\bibinfo {author} {\bibfnamefont {T.}~\bibnamefont
  {Betz}}, \bibinfo {author} {\bibfnamefont {D.}~\bibnamefont {Koch}}, \bibinfo
  {author} {\bibfnamefont {Y.-B.}\ \bibnamefont {Lu}}, \bibinfo {author}
  {\bibfnamefont {K.}~\bibnamefont {Franze}},\ and\ \bibinfo {author}
  {\bibfnamefont {J.~A.}\ \bibnamefont {K{\"a}s}},\ }\bibfield  {title}
  {\bibinfo {title} {Growth cones as soft and weak force generators},\ }\href
  {https://doi.org/10.1073/pnas.1106145108} {\bibfield  {journal} {\bibinfo
  {journal} {Proc. Nat. Acad. Sci. U.S.A.}\ }\textbf {\bibinfo {volume}
  {108}},\ \bibinfo {pages} {13420} (\bibinfo {year} {2011})},\ \Eprint
  {https://arxiv.org/abs/10.1073/pnas.1106145108} {10.1073/pnas.1106145108}
  \BibitemShut {NoStop}%
\end{thebibliography}%

% or Copy of BibTeX .bbl output:

% \begin{thebibliography} etc...

% \end{thebibliography}%

% End of BibTeX .bbl output.

\end{document}